\documentclass[
 reprint,
 superscriptaddress,
 amsmath,amssymb,
 aps,
]{revtex4-2}

\usepackage{graphicx}
\usepackage{comment}
\usepackage{float}
\usepackage[caption=false]{subfig}
\usepackage{dcolumn}
\usepackage{bm}
\usepackage[table]{xcolor} 
\usepackage{tabularx}
\usepackage{booktabs}
\usepackage{multirow}
\usepackage{array}    
\usepackage[
    colorlinks=true,
    linkcolor=blue,    
    citecolor=blue,    
    urlcolor=black      
]{hyperref}

\begin{document}

\preprint{APS/123-QED}

\title{Gravitational Wave Signal Extraction Against Non-Stationary Instrumental Noises with Deep Neural Network}

\author{Yuxiang Xu}
\affiliation{Hangzhou Institute for Advanced Study, UCAS, Hangzhou 310024, China}
\affiliation{Center for Gravitational Wave Experiment, National Microgravity Laboratory, Institute of Mechanics, Chinese Academy of Sciences, Beijing 100190, China}
\affiliation{Shanghai Institute of Optics and Fine Mechanics, Chinese Academy of Sciences, Shanghai 201800, China}
\affiliation{Taiji Laboratory for Gravitational Wave Universe (Beijing/Hangzhou), University of Chinese Academy of Sciences (UCAS), Beijing 100049, China}
\author{Minghui Du}
\affiliation{Center for Gravitational Wave Experiment, National Microgravity Laboratory, Institute of Mechanics, Chinese Academy of Sciences, Beijing 100190, China}
\author{Peng Xu}
\email{Corresponding author: xupeng@imech.ac.cn}
\affiliation{Center for Gravitational Wave Experiment, National Microgravity Laboratory, Institute of Mechanics, Chinese Academy of Sciences, Beijing 100190, China}
\affiliation{Hangzhou Institute for Advanced Study, UCAS, Hangzhou 310024, China}
\affiliation{Lanzhou Center of Theoretical Physics, Lanzhou University, Lanzhou 730000, China}
\affiliation{Taiji Laboratory for Gravitational Wave Universe (Beijing/Hangzhou), University of Chinese Academy of Sciences (UCAS), Beijing 100049, China}
\author{Bo Liang}
\affiliation{Hangzhou Institute for Advanced Study, UCAS, Hangzhou 310024, China}
\affiliation{Center for Gravitational Wave Experiment, National Microgravity Laboratory, Institute of Mechanics, Chinese Academy of Sciences, Beijing 100190, China}
\affiliation{Shanghai Institute of Optics and Fine Mechanics, Chinese Academy of Sciences, Shanghai 201800, China}
\affiliation{Taiji Laboratory for Gravitational Wave Universe (Beijing/Hangzhou), University of Chinese Academy of Sciences (UCAS), Beijing 100049, China}
\author{He Wang}
\affiliation{CAS Key Laboratory of Theoretical Physics, Institute of Theoretical Physics, Chinese Academy of Sciences, Beijing 100190, China}
\affiliation{International Centre for Theoretical Physics Asia-Pacific, University of Chinese Academy of Sciences, 100190 Beijing, China}
\affiliation{Taiji Laboratory for Gravitational Wave Universe (Beijing/Hangzhou), University of Chinese Academy of Sciences (UCAS), Beijing 100049, China}

\date{\today}

\begin{abstract}
Sapce-borne gravitational wave antennas, such as LISA and LISA-like mission (Taiji and Tianqin), will offer novel perspectives for exploring our Universe while introduce new challenges, especially in data analysis. 
Aside from the known challenges like high parameter space dimension, superposition of large number of signals etc., gravitational wave detections in space would be more seriously affected by anomalies or non-stationarities in the science measurements.
Considering the three types of foreseeable non-stationarities including data gaps, transients (glitches), and time-varying noise auto-correlations, which may come from routine maintenance or unexpected disturbances during science operations, we developed a deep learning model for accurate signal extractions  confronted with such anomalous scenarios. 
Our model exhibits the same performance as the current state-of-the-art models do for the ideal and anomaly free scenario, while shows remarkable adaptability in extractions of coalescing massive black hole binary signal against all three types of non-stationarities and even their mixtures.
This also provide new explorations into the robustness studies of deep learning models for data processing in space-borne gravitational wave missions.

\end{abstract}

\maketitle

\section{\label{sec:introduction}INTRODUCTION}

Since the first landmark event GW150914 captured by Adv-LIGO \cite{abbott2016gw150914},  today nearly a hundred Gravitational Wave (GW) signals from mergers of stellar mass compact binaries had been observed by the LIGO-Virgo collaboration, which lead to numerous accomplishments in astrophysics and fundamental physics \cite{abbott2016observation,abbott2016gw151226,abbott2016binary,scientific2017gw170104,abbott2017gw170608,abbott2017gw170814,abbott2017gw170817,abbott2020gw190425,abbott2020gw190412,abbott2020gw190521,abbott2021tests,ezquiaga2021hearing,bozzola2021general}.
However, limited by seismic noises, ground-based GW detectors are typically sensitive to signals at frequencies higher than 10 Hz \cite{abbott2019gwtc,abbott2020prospects,freise2010interferometer}. 
To enclose the exciting sources of larger and heavier astrophysical systems, one needs to explore the low frequency band of the GW spectrum with detectors of much longer baselines.
The first generation space-borne antennas, including LISA (Laser Interferometer Space Antenna) \cite{amaro2017laser,baker2019laser} and the LISA-like missions Taiji \cite{gong2011scientific,hu2017taiji} and Tianqin \cite{luo2016tianqin}, would be launched in the 2030s and cover the mHz band.
Offering novel perspectives for exploring the Universe, such space-borne antennas will also introduce new challenges \cite{baker2019laser}, especially in data analysis \cite{dey2021effect,baghi2022detection,edwards2020identifying}.

Space antennas are supposed to response to a large amount of superimposed GW signals, including coalescing (super) Massive Black Hole Binaries (MBHB), extreme mass ratio inspirals, galactic compact binaries, evolving topological defects, primordial GW background, and also un-modeled sources. 
To achieve the expected scientific objectives, throughout analysis of the observational data and estimations of the relevant parameters from the largely superimposed GW signals against  complicated noises are needed. 
Many works have been done to address these issues \cite{mock2006overview,katz2022fully,vallisneri2005synthetic},  and in present days the most accurate results can be obtained by matched filtering method for signals buried in Gaussian noises \cite{finn1992detection,usman2016pycbc}. 
However, the construction of a full-fledged pipeline of scientific data analysis for space-borne antennas remains still an unfinished task, especially with the boost, in recent years, from machine learning methods been included.
Following the studies  in \cite{zhao2023space,wang2020gravitational},  accurate signal waveform extractions can be viewed as an intermediate step of the subsequent high-precision parameter estimations with machine learning methods, and the improvements in its performances and robustness are worth further explorations and investigations.


As lessons learned from the LIGO-Virgo observatories \cite{de2020primordial} and precedent missions like GRACE/GFO \cite{chen2022applications}, LISA PathFinder  (LPF) \cite{armano2009lisa} and also Taiji-1 \cite{taiji2021taiji}, new issues in data analysis for space-borne GW detections begin to draw more attentions \cite{baghi2019gravitational,armano2018beyond,robson2019detecting,armano2022transient,edwards2020identifying}. 
As mentioned, to precisely measure the related parameters and infer the physical properties of the sources, continues measurements without disruptions in the data are normally demanded, which however imposes hard challenges on the long-term stability and robustness of both the payloads and satellite platforms.
GW detection in space would be more seriously affected by anomalies or non-stationarities in the science measurements, including instrument transients or noise bursts, data gaps, slowly varying noise auto-correlations, trends etc., as the detailed knowledge of instrumental noises is crucial to GW signals extraction and parameter estimations. 
According to the designs of the LISA and Taiji missions, the foreseeable anomalies or non-stationarities in science measurements may come from the routine maintenance of the satellites or the onboard payloads and unexpected environment disturbances or instrumental anomalies, which will be the main subjects of this work. 
Cyclostationary properties also exist mainly due to the orbital modulations of the stochastic GW foreground from compact binaries, which are not within the scope of this work and will be left for future studies.

Routine maintenance, such as antenna re-orientations, payloads calibrations etc., could cause important and even continuous disturbances of satellite platforms and affect heavily the performances of the key payloads, therefore result into varying noise auto-correlations and even data loss or gaps.
For example, the scheduled re-orientations of the transmission antennas would lead to data gaps about 3.5 hours per week or 7 hours every two-weeks  \cite{dey2021effect}.
In addition, unforeseen anomalies or failures due to hardware problems would take place and produce random unscheduled data gaps.  
Based on the experiences from LPF \cite{armano2016sub,armano2018beyond,baghi2022detection,armano2022transient},  GRACE/GFO \cite{frommknecht2007integrated,sheard2012intersatellite} and also Taiji-1 \cite{taiji2021china}, instrumental transients may also contaminate the science data, such as the acceleration glitches in gravitational reference sensor (GRS) systems, phase jumps in interferometers etc. 
Early studies on data processing algorithms against such possible anomalies can be found in \cite{baghi2019gravitational,robson2019detecting,armano2022transient}.
Recently, Dey et al. \cite{dey2021effect} have used Bayesian analysis techniques to assess the impact of data gaps and concluded that the effects of unscheduled gaps were often more significant, and Spadaro et al. \cite{spadaro2023glitch} used a joint parameter estimation approach to evaluate the influences of GRS glitches demonstrating that it is possible to accurately identify glitches in the absence of overlapping GW signals. 
These studies treated the above non-stationarities independently, and, for instrumental transients, prior knowledge of transient models are required.
In this work, given the potentials and capacities of deep neural networks, we seek to develop an anomaly-model-independent signal extraction method for space GW antennas that can thoroughly treat and overcome the impacts of the above three types of data non-stationarities or anomalies.


Deep learning methods have already achieved considerable success in data processing of GW detections \cite{george2018deep,gabbard2018matching,wang2020gravitational,gabbard2022bayesian,dax2021real,colgan2020efficient,cavaglia2019finding,razzano2018image,ormiston2020noise,torres2016denoising,wei2020gravitational,zhao2023space}. 
Compared to traditional algorithms, such as matched filtering \cite{finn1992detection,usman2016pycbc,cannon2021gstlal}, which generally require a vast template bank and thus consume substantial computational resources, deep learning methods are well known for their fast processing speed and excellent generalization ability, making it possible to achieve rapid and anomaly-model-independent signal extraction. 
For GW detection in space, Conv-TasNet \cite{zhao2023space} has demonstrated its performance in various scenarios,   {completing signal extraction in $\lesssim$ \( 10^{-2} \) seconds}, however, robustness tests under anomalous conditions are overlooked.
In this work, focusing on the three types of possible non-stationarities for LISA or LISA-like missions, including gaps, transients (glitches), and time varying noise auto-correlations, we developed a deep learning model for accurate extractions of coalescing MBHB signals confronted with such anomalous scenarios.
This also provide new explorations into the robustness studies of deep learning models for data processing in space-borne GW missions.
Our model exhibits the same performance as the Conv-TasNet model \cite{zhao2023space} does for the ideal and anomaly free scenario, while shows remarkable adaptability in signal extractions against all three types of non-stationarities and even their mixtures.

\section{\label{sec:model}Dense-LSTM model}

\begin{figure*}[hbt]
\includegraphics[width=0.9\textwidth]{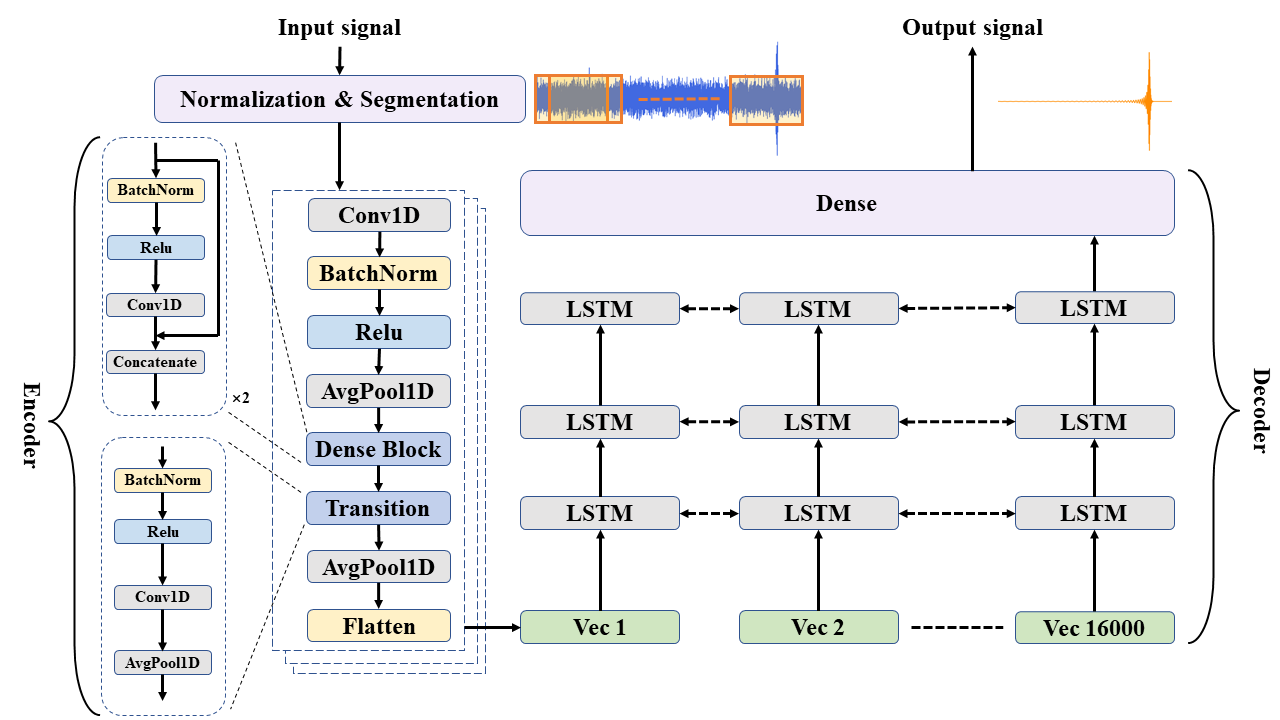}
\caption{\label{fig1}The architecture of our denoising autoencoder is delineated as follows. The input data is first subjected to normalization and segmentation (top left), resulting in the formation of overlapping subsequences. Each subsequence is then processed through the encoder to extract its characteristic feature vector. This is followed by passage through three bidirectional LSTM layers to yield predictive values. The final reconstructed waveform is then attained in the Dense layer (top right).}
\end{figure*}

Autoencoders, commonly employed as denoising architectures, are extensively applied in fields like image denoising \cite{gondara2016medical,xie2012image,vincent2010stacked,creswell2018denoising}, signal extractions \cite{chiang2019noise,saad2020deep,xia2017intelligent,dasan2021novel}, and speech enhancements, et al. \cite{araki2015exploring,lai2016deep,feng2014speech,lu2013speech}. 
Autoencoder models include an encoder and a decoder.
The encoder is designed to extract feature mappings from the input data, while the decoder is to generate reconstructed data from these mappings.
For GW signal extractions in LIGO-Virgo data analysis, the CNN-LSTM denoising autoencoder model has been used \cite{chatterjee2021extraction} and shown superior performance.

As discussed in Sec. \ref{sec:introduction}, the long periods and consequently the much more complicated data anomalies for science measurements of LISA and LISA-like missions have driven us to develop models capable of uncovering deeper mappings within the signals and thereby reducing the loss and alteration of the crucial information under significant anomalous conditions. 
Simple CNN architectures fall short in meeting these demands. 
In this work,  we choose to use a deeper CNN network different from the CNN-LSTM denoising autoencoder suggested by Chatterjee et al. \cite{chatterjee2021extraction}. 
The DenseNet is chosen as the encoder for its advantage in reducing computational overhead while maintaining comparable accuracy compared with other deep CNN networks \cite{krizhevsky2012imagenet,simonyan2014very,he2016deep,lecun2015deep} and at the same time ensuring faster feature extraction. 
More importantly, DenseNet thoroughly implements the concept of feature reuse, with each layer directly connected to all preceding layers. 
This enables the comprehensive utilization of low-complexity shallow features, facilitating the derivation of a smoother and more robust decision function. 
Such a design ensures considerable extraction capabilities even when confronted with data strongly contaminated by the possible non-stationarities or anomalies for space-borne GW detections. 
For the decoder, we retain the bidirectional LSTM network due to its superiority in capturing long-term dependencies within sequences.

The structure of our model is depicted in Fig~\ref{fig1}. 
The input GW data for the model have a sampling frequency of 0.1Hz and a duration of 160,000 seconds, amounting to 16,000 data points. 
The detailed data preparations can be found in Sec~\ref{sec:datasets}. 
All data are normalized between -1 and 1, followed by segmentation into 16,000 overlapping subsequences of length 4. 
Each subsequence is then fed into the decoder network for feature extraction. 
The feature vector of each subsequence is processed through three bidirectional LSTM layers for predicting the output of the subsequent time step. 
Finally, a complete waveform output is obtained through a dense layer.

\section{\label{sec:datasets}Data preparations}

For LISA and LISA-like missions, laser frequency instability noise is dominant in the science measurements, and a pre-processing technology called Time Delay Interferometry (TDI) \cite{Tinto,armstrong1999time,Babak2021}  is employed to efficiently suppress such noise.
Following this, our data is generated through the first generation noise orthogonal TDI channels \( A \) and \( E \), see \cite{Tinto} for a detailed introduction. {  The total TDI output data stream of channel $I$ (dubbed $s^I(t)$) is the combination of signal $h^{I}(t)$ and noise $n^{I}(t)$:
\begin{equation}
    s^{I}(t) = h^{I}(t) + n^{I}(t), \quad I \in \{A, E\},  
\end{equation}
where the detector response in terms of  channel $I$ to some incident GW with polarizations $h_\alpha$ reads
\begin{equation}
h^I(t) = \sum_{\alpha} T_{\alpha}^{I}(t) \ h_{\alpha}(t), \quad \alpha \in \{+, \times\}, 
\label{eq:aef}
\end{equation}
where \( T_{\alpha}^{I}(t) \) is the total  transfer function of signal including antenna responses and TDI combinations, and \( h_{\alpha}(t) \) represents the polarization components of the coalescing MBHB GW signal.}
Specific details of this transfer function can be found in \cite{Babak2021,Tinto}.   {In this paper, the TDI transfer functions of signals are derived according to  Taiji's mission concept and orbit configurations \cite{luo2020brief}, which has a nominal arm length of $\sim 3\times 10^9$ m.}
The PyCBC package \cite{nitz2022gwastro} and the waveform model SEOBNRv4 \cite{bohe2017improved} are employed to generate the coalescing MBHB GW templates $h_{\alpha}(t)$. 
{  On the other hand, }
the noise floor of the output data from a certain TDI channel is in principle determined by the corresponding TDI combinations of GRS residual acceleration noises $n_{\rm ACC}(t)$ and optical metrology system noises $n_{\rm OMS}(t)$.  
{  The resulting total instrumental noise of TDI channel $I$ can be written as 
\begin{equation}\label{eq:noise_transfer_function}
    n^I(t) = T^{I}_{\rm OMS}(t)n_{\rm OMS}(t) + T^{I}_{\rm ACC}(t)n_{\rm ACC}(t),
\end{equation}
with $T^I_{\rm OMS}(t)$ and $T^I_{\rm ACC}(t)$ being the transfer functions of GRS residual acceleration noise and optical metrology system noise, respectively~\cite{Tinto,Babak2021}, which are common to LISA and Taiji missions. 
The time series of basic instrumental noises $n_{\rm ACC}(t)$ and $n_{\rm OMS}(t)$ are generated according to the Power Spectral Densities (PSDs) of these two noise components, denoted as $S_{\text{OMS}}^{1/2}(f)$ and $S_{\text{ACC}}^{1/2}(f)$, 
 and according to the designs of Taiji \cite{luo2020brief} we assume the nominal values
\begin{equation}
S_{\text{OMS}}^{1/2}(f) = 8 \times 10^{-12} \sqrt{1 + \left(\frac{2\text{mHz}}{f}\right)^4} \, \frac{\rm m}{\sqrt{\rm Hz}},\label{eq:SOMS}
\end{equation}
\begin{equation}
\begin{split}
S_{\text{ACC}}^{1/2}(f) = & 3 \times 10^{-15} \sqrt{1 + \left(\frac{0.4\text{mHz}}{f}\right)^2} \\
& \times \sqrt{1 + \left(\frac{f}{8\text{mHz}}\right)^4} \, \frac{\rm m/s^2}{\rm \sqrt{Hz}}.
\label{eq:SACC}
\end{split}
\end{equation}}
\begin{table}[t]
\caption{\label{tab:mbhb_space}Summary of parameter setups in coalescing MBHB GW signals generations.}
\begin{ruledtabular}
\renewcommand{\arraystretch}{1.2}
\begin{tabular}{lcc}
Parameter & Lower bound & Upper bound\\
\colrule
\( M_{\text{tot}} \) \footnote{Total mass of the system.} & \( 10^6 M_{\odot} \) & \( 10^8 M_{\odot} \) \\
\( q \) \footnote{Mass ratio of the objects in the binary system.} & 0.01 & 1 \\
\( s_1^z \) \footnote{Dimensionless spin parameter for the primary object.} & -0.99 & 0.99 \\
\( s_2^z \) \footnote{Dimensionless spin parameter for the secondary object.} & -0.99 & 0.99 \\
\end{tabular}
\end{ruledtabular}
\end{table}
\begin{figure}[b]
\centering
\includegraphics[width=0.48\textwidth]{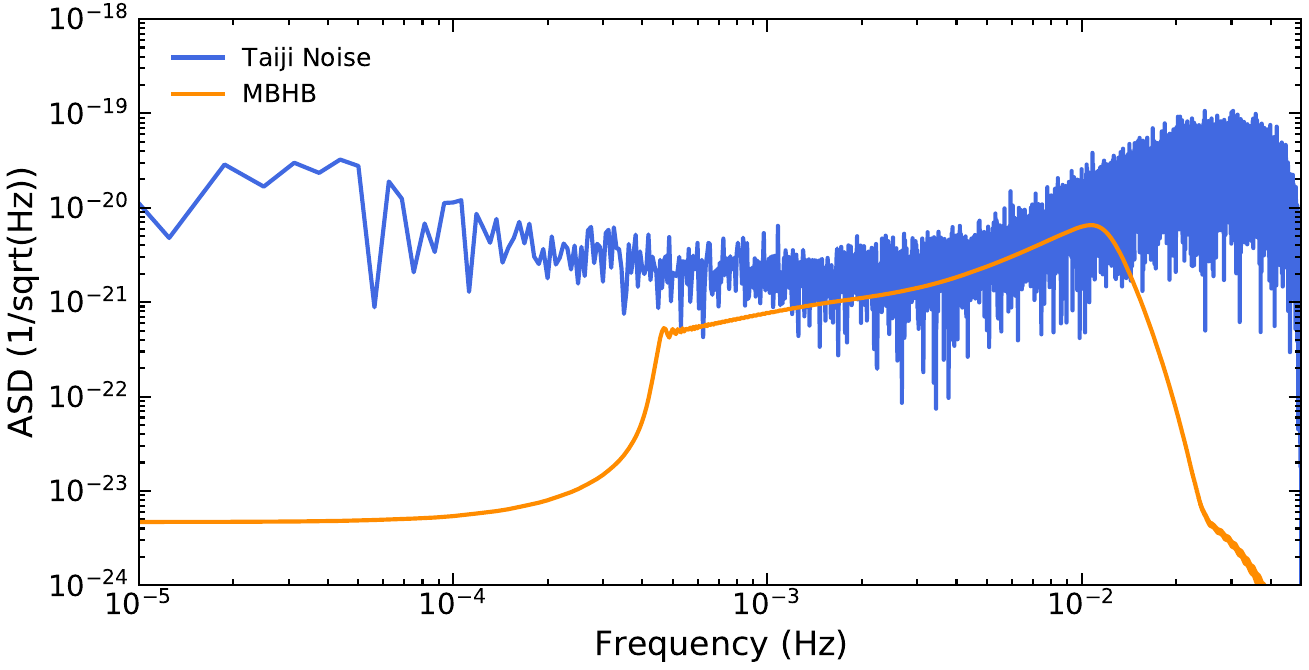}
\caption{\label{fig:data}The noise curve and the typical coalescing MBHB signal (SNR$\sim$ 35.0) in the frequency domain.}
\end{figure}
  {Specifically, to generate the noise data in time domain according to the PSDs, we convert Gaussian white noises to the frequency domain through Fourier transform, then multiply them by the amplitude spectral density value at each frequency to adjust the spectra of the noises, and finally obtain  time series of noises through inverse Fourier transform.
Theoretically, the total noise  PSDs of the \( A \), \( E \) channels \( S_{n}^{I}(f),\  I \in \{A,  E\} \) are the TDI combinations of  $S_{\rm OMS}(f)$ and $S_{\rm ACC}(f)$ calculated according to Eq.~(\ref{eq:noise_transfer_function}),  
see Refs.~\cite{Babak2021,Tinto,wang2020numerical,wang2023revisiting} for their specific  forms. }

We established the parameter space for coalescing MBHBs as in Table~\ref{tab:mbhb_space}, and employed a uniform grid to generate the diverse waveforms. 
Parameters such as times and phases at coalescence, sky locations, polarization angles, as well as inclinations are drawn from uniform distributions. 
We inject the signal with specific optimal Signal-to-Noise Ratios (SNR)
\begin{equation}
{\rm SNR} = (h^I | h^I)^{1/2},
\label{eq:snr}
\end{equation}
where \( h^I \) represents the signal template through the TDI channel $I$, the inner product \( (s^I | h^I) \) is defined as,
\begin{equation}
(s^I | h^I) = 2 \int_{f_{\text{min}}}^{f_{\text{max}}} \left(\tilde{s}^I(f) \tilde{h}^{I*}(f) + \tilde{s}^{I*}(f) \tilde{h}^I(f)\right) \, df,
\label{eq:inner_product}
\end{equation}
here, \( f_{\text{min}} = 3 \times 10^{-5} \) Hz and \( f_{\text{max}} = 0.05 \) Hz.
The tilde symbol represents the Fourier transform and \( * \) the complex conjugation.   {This inner product can also be used to calculate the overlap between the signal waveform $o$ extracted by the model and the whitened template waveform $h$ through channel $A$ or $E$ can be evaluated in terms of the overlap function
\begin{equation}
\mathcal{O}(o, h) = \max_{t_c, \phi_c} (\hat{o}|\hat{h}),
\end{equation}
where \( \hat{o} \) and \( \hat{h} \) are the normalized ones, and \( t_c \) and \( \phi_c \) are the instantaneous time and phase corresponding to the maximum overlap between  \( \hat{o} \) and \( \hat{h} \).}

We set the SNR level for the training dataset as 50, and generate 10,000 samples.
The distances of the sources are adjusted according to the specific optimal SNR.
All samples are subjected to whitening processing, and a Tukey window with \(\alpha = \frac{1}{8}\) is adopted.

Following the same routine, we produce 2,000 test samples, see Fig~\ref{fig:data} for illustration. 
In order to assess the model's performance for signals extractions with relative low SNRs and test the robustness of our model against various data anomalies, these samples were generated with SNRs ranging from 30 to 70.
\begin{figure}[hbt]
\centering
\includegraphics[width=0.48\textwidth]{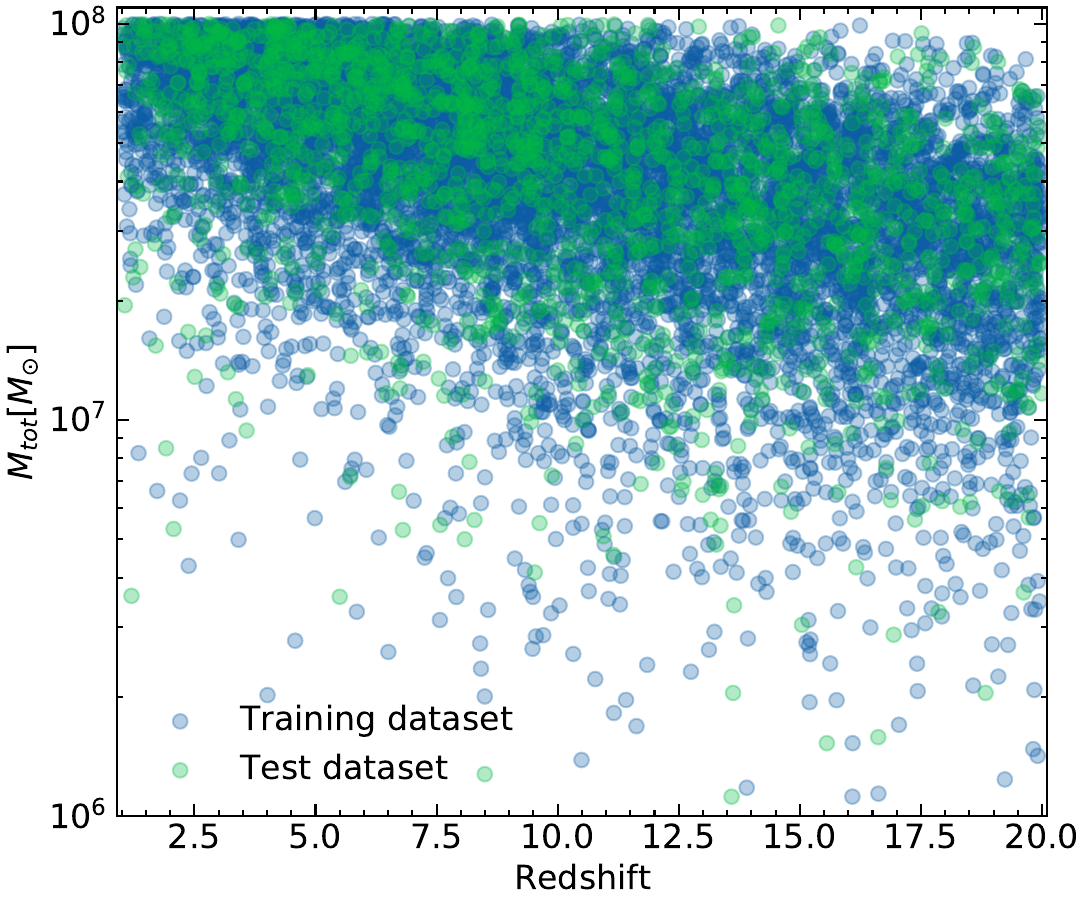}
\caption{\label{fig:mbhb_dis}The figure illustrates the distribution of our training and test datasets across redshifts ranging from 1 to 20.}
\end{figure}
  {Our training and test datasets were sampled from redshifts ranging from 1 to 20. The resulting distribution is shown in Fig~\ref{fig:mbhb_dis}.}

\section{\label{sec:strategy}Training strategy}

For denoising autoencoders, the Mean Squared Error (MSE) is a prevalent loss function employed to quantify the discrepancy between the network's predictions and the true values. 
However, given that the GW signals fed into the network have undergone normalization, the contributions to the overall MSE from the earlier stages of the coalescing signals are rather weak compared to those from the merger phases. 
This can hinder the model from achieving optimal refinements. 
To address this, the loss function defined in \cite{chatterjee2021extraction} is adopted which is the difference of the MSE and the fractal Tanimoto similarity coefficient \cite{diakogiannis2020looking} terms
\begin{equation}
L_{o,h} = \frac{\sum_{i}^{n}(h_i - o_i)^2}{n} - r^d_{w,o,h}, \quad (i=1,\ldots,n)
\end{equation}
the first term on the right hand side is the MSE term and the fractal Tanimoto similarity coefficient is defined as
\begin{equation}
r^{d}_{w,o,h} = \frac{\sum_{i}^{n} w_{i} \cdot o_{i} \cdot h_{i}}{2^{d} \sum_{i}^{n} w_{i} \cdot (o_{i}^{2} + h_{i}^{2}) - (2^{d+1} - 1) \sum_{i}^{n} w_{i} \cdot o_{i} \cdot h_{i}}.
\end{equation}
The MSE is used for the optimizations of individual data points, while the fractal Tanimoto similarity coefficient is used for the optimizations of the overall data. 
Tanimoto coefficient is commonly used to measure the similarity between two vectors, and the fractal coefficient, as an enhanced version, introduces a parameter \(d\) to facilitate deeper levels of optimizations. 
This approach ensures that the extracted data achieves good amplitude and phase agreements with the template data on a global level.

Another worth-noting modification is that, unlike in the original work \cite{chatterjee2021extraction} where the weight \( w_{i} \) is set to \( 1/o_{i} \), we set \( w_{i} \) equal to 1 in our approach. 
This is because, the long observation time in space borne GW detection
will give rise to large number of data points that having amplitudes close to zero in the generated templates, the high weights, such as $\omega_i = 1/o_i$, of these data points with $o_i\sim 0$ will reduce the weights of the data from the merger phases, and therefore the optimizations will be highly affected by the data segments away from the important merge phases and consequently prevent the network from converging. For the parameter \( d \), which is used to adjust the depth of optimization in similarity, is initially set to 0. 
As the training converges, \( d \) grows gradually to facilitate deeper levels of optimization. 

To summarize, at the beginning of the training, we set the learning rate to \(10^{-3}\) and \(d\) to 0. 
As the model approaches convergence, we set \(d=5\) and reduce the learning rate by a factor of 10. 
We have chosen the learning rate as \(10^{-3}\), \(10^{-4}\), and \(10^{-5}\) with corresponding \(d\) taking values as 0, 5, and 10 for case studies, and trained for 200 epochs for each case. 

\section{\label{sec:results}Results}

\begin{figure}[h]
\centering
\includegraphics[width=0.48\textwidth]{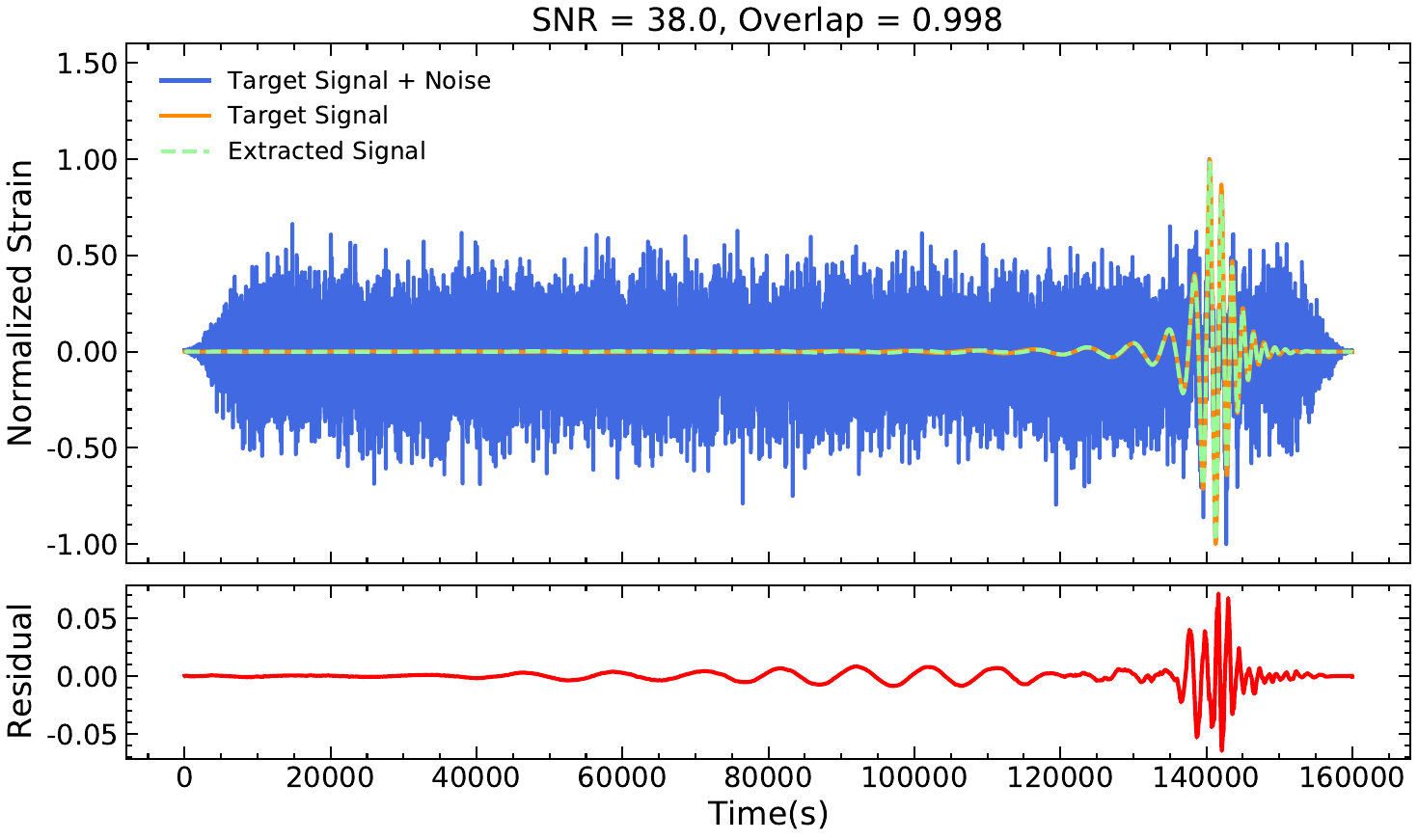}
\caption{\label{fig:norm}Example of signal extraction with our model when there is no anomaly or non-stationarity present in the data.}
\end{figure}

We first tested the extraction performance of our model for the ideal case without data non-stationarities or anomalies, see illustrations in Fig~\ref{fig:norm}. 
More than 99\% of the signals extracted by the model achieve an overlap $\geq$ 0.9 with the corresponding template, and each signal could be extracted in less than \(10^{-2}\) seconds. 
This demonstrates a similar performance of our model compared to the current state-of-the-art models \cite{zhao2023space}. 
In the followings, we conduct various tests of our model against the possible non-stationarities or anomalies known to the literature, including data gaps, time-varying noise auto-correlations, glitches, and even their mixtures. 
We expect that at least 80\% of the signals extracted by the model should attain an overlap $\geq$ 0.9 with the template signals, which we will use as the criterion for assessing the limits of the model's performance.
The physical meanings of a more realistic criterion or threshold should come from more detailed studies on false positive, uncertainties in parameters estimations and so on, which will be left for future studies. 

\subsection{\label{sec:gaps}Data gaps}

\begin{figure*}[htb]
\centering
\includegraphics[width=0.98\textwidth]{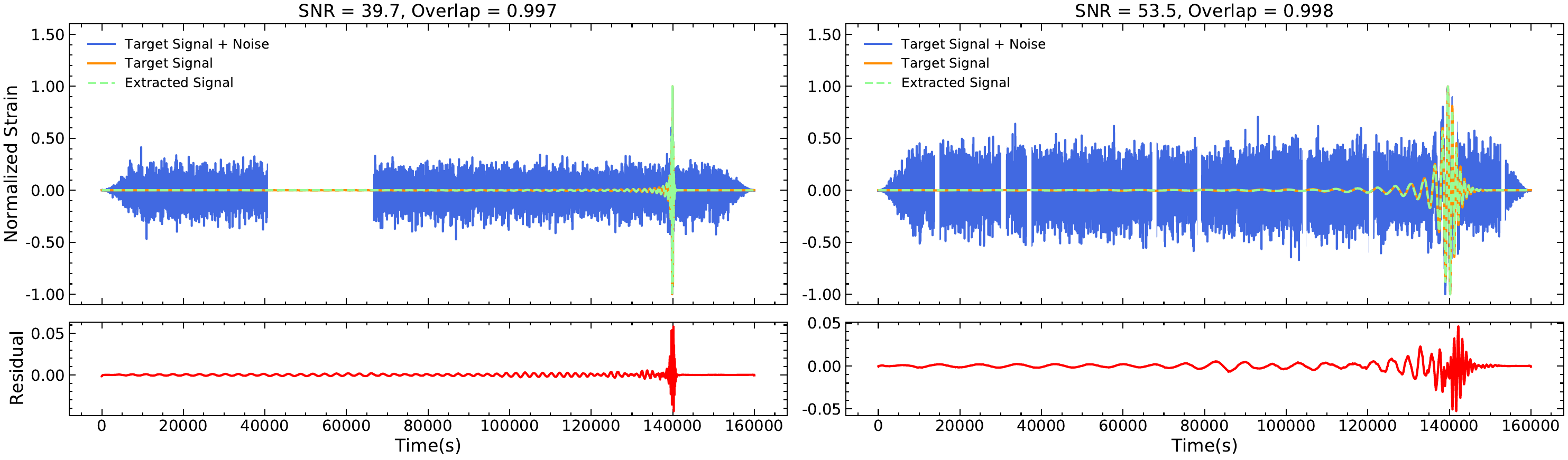}
\caption{\label{fig:gap1}The left image illustrates the model's extraction result in the presence of scheduled long gap, while the right image depicts the outcomes when facing unscheduled gaps, with a occurrence frequency about once every four hours.}
\end{figure*}
\begin{figure*}[htb]
\centering
\includegraphics[width=0.98\textwidth]{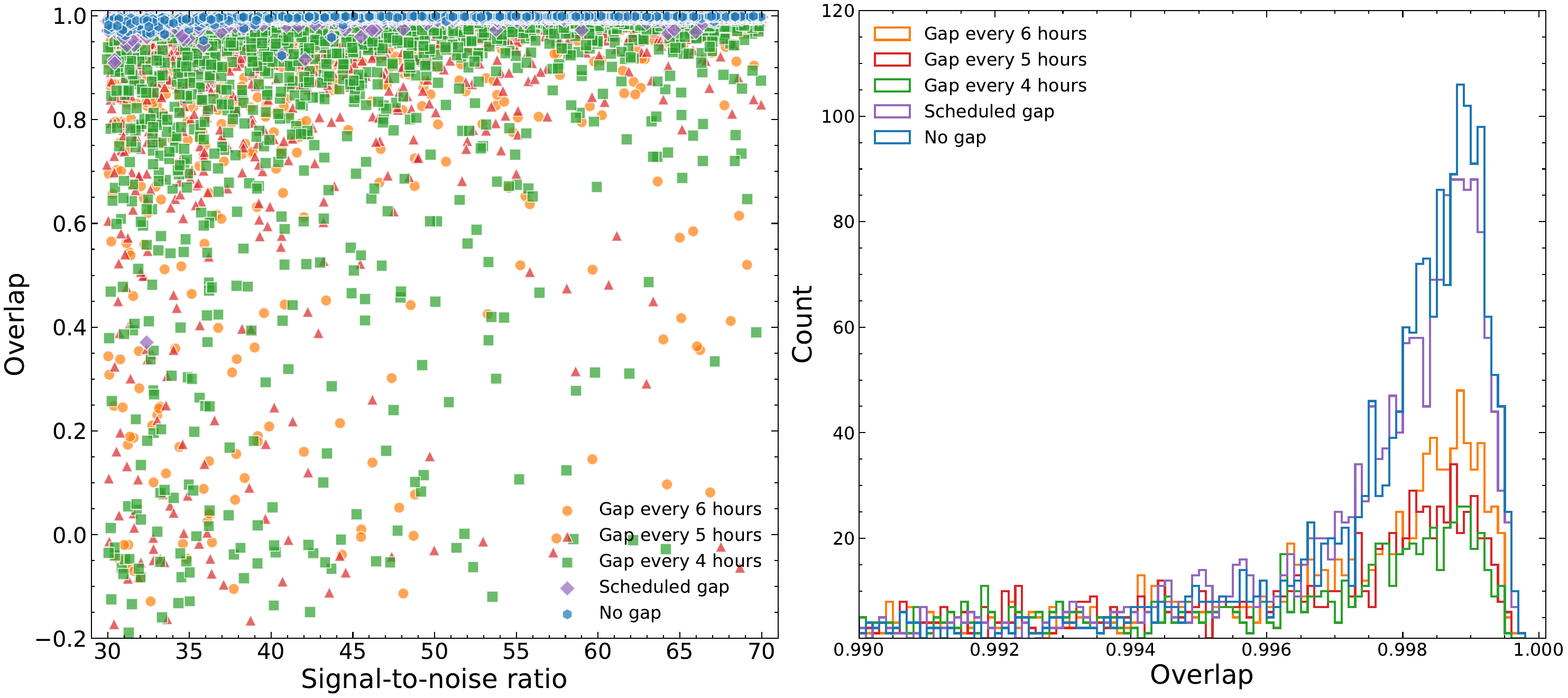}
\caption{\label{fig:gap2}The left image displays the distributions of the overlaps for our test set confronting with data gaps of different occurrence frequencies. The image on the right presents the histogram for instances where the model achieves an overlap $\geq$ 0.99, corresponding to our test set confronting with data gaps of different occurrence frequencies.}
\end{figure*}

Following the discussions in \cite{dey2021effect},  we use window functions to generate data gaps (see Ref~\cite{dey2021effect} for the specific forms), which, as discussed in Sec. \ref{sec:introduction} can be classified into two types, that the scheduled prolonged gaps and unscheduled random but short gaps. 
For the scheduled gaps, we emulate an extreme situation, resulting into a data loss of 7 hours during the inspiral phases. 
For unscheduled gaps, it can last up to days or cause data loss of a few minutes per day due to glitch masking according to \cite{baghi2019gravitational}. Considering the length of our data, here we choose the second type of unscheduled gap.
We set each gap to cause 5 to 8 minutes of data loss, and progressively increase the frequency of occurrences of such randomly happened gaps to a threshold at which our model can hardly tolerate. 
Moreover, the unscheduled gaps are allowed to break out at any place of our data, while a prolonged data gap during the merger phase is ignored for obvious reason. 
Our extraction results are presented in Fig~\ref{fig:gap1}. 
Our model can proficiently extract the signals despite facing these two types of data gaps. 

We delved further into the threshold performance of our model, as detailed in Fig~\ref{fig:gap2}. 
The model exhibits minimal accuracy loss in scenarios with scheduled gaps before mergers. 
When confronted with the random unscheduled gaps, the model's results gradually worsen as the frequency of occurrences increases.
As expected, signals with lower SNRs are more susceptible to data gaps. 
\begin{table}[htb]
\caption{\label{tab:gap}For different types of data gaps, the ratio of overlaps $\geq$ 0.9 between the model-extracted signals and the templates across the entire dataset. The last two columns represent the means and standard deviations of the overlap.}
\begin{ruledtabular}
\renewcommand{\arraystretch}{1.2}
\begin{tabular}{lccc}
Type of gap & Overlaps $\geq$ 0.9 & Mean & Std. Dev. \\
\colrule
No gap & 100.0\% & 0.997 & 0.005 \\
Scheduled gap & 99.9\% & 0.996 & 0.016 \\
Gap every 6 hours & 86.7\% & 0.935 & 0.169 \\
Gap every 5 hours & 83.1\% & 0.923 & 0.181 \\
Gap every 4 hours & 77.3\% & 0.893 & 0.220 \\
\end{tabular}
\end{ruledtabular}
\end{table}
\begin{figure*}[htb]
\centering
\includegraphics[width=0.98\textwidth]{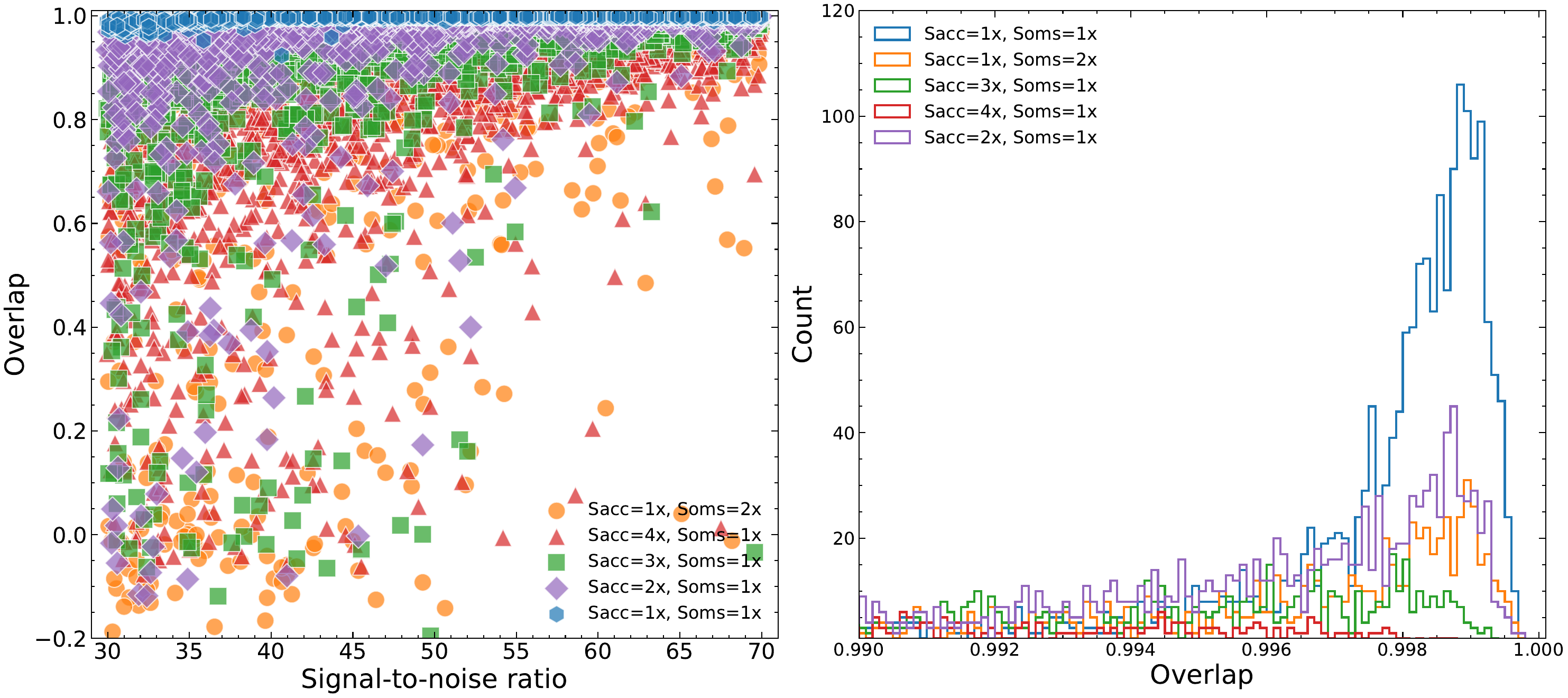}
\caption{\label{fig:noise2}The left image displays the distributions of the overlaps for our test set confronting with time-varying noise auto-correlations of varying intensities. The image on the right presents the histogram for instances where the model achieves an overlap $\geq$ 0.99, corresponding to our test set confronting with time-varying noise auto-correlations of varying intensities.}
\end{figure*}
In Tab~\ref{tab:gap}, we present more detailed results under different conditions. 
According to the predetermined criterion, our model can maintain robustness even when the frequency of unscheduled gaps reaches once per 5-hours.

\subsection{\label{sec:acc}Time-varying noise auto-correlations }

\begin{figure}[b]
\centering
\includegraphics[width=0.48\textwidth]{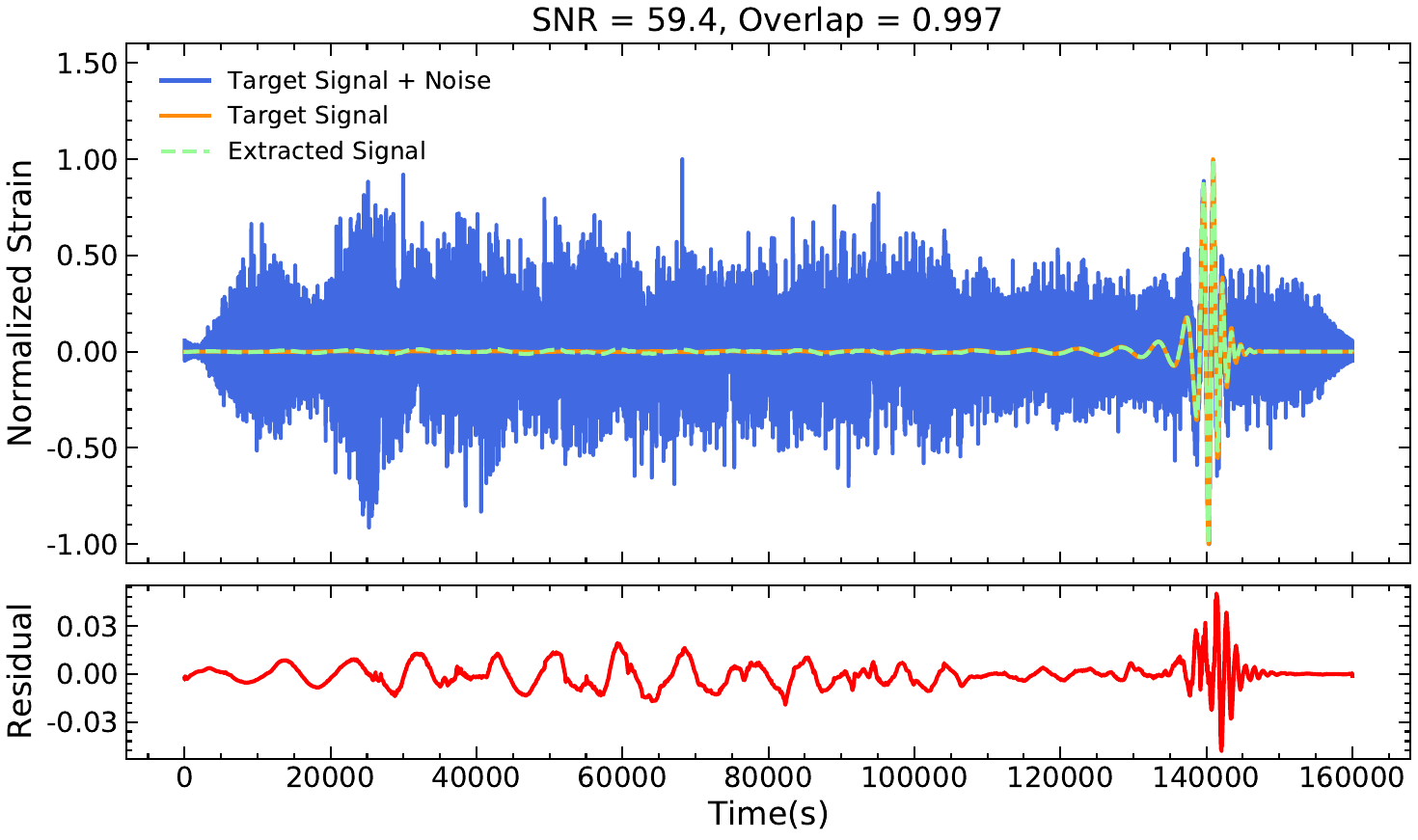}
\caption{\label{fig:noise1}Example of signal extraction of the model when contending with time-varying noise  auto-correlations. Here, a segment of the data spanning 80,000 seconds has the residual acceleration noise set to threefold its nominal magnitude. The result demonstrate that our model effectively extracted the signal, achieving a high overlap.}
\end{figure}

We simulated non-stationary noises with time-varying auto-correlations or PSDs by altering the magnitudes of \( S_{\text{OMS}}(f) \) and \( S_{\text{ACC}}(f) \) of the two noise components, with the modified noise segment lasting for 80,000 seconds. 
In Fig~\ref{fig:noise1}, we present one of our extraction samples. 
The limits of our model's performance against such non-stationary noise is investigated and shown in Fig~\ref{fig:noise2}. 
As expected, data with lower SNRs will be more significantly affected by time-varying noise PSDs, and the detailed results are presented in Tab~\ref{tab:noise}. 
Our model can successfully extract the signal without being affected by such variations in total noise PSDs.

\begin{figure}[h]
\centering
\includegraphics[width=0.48\textwidth]{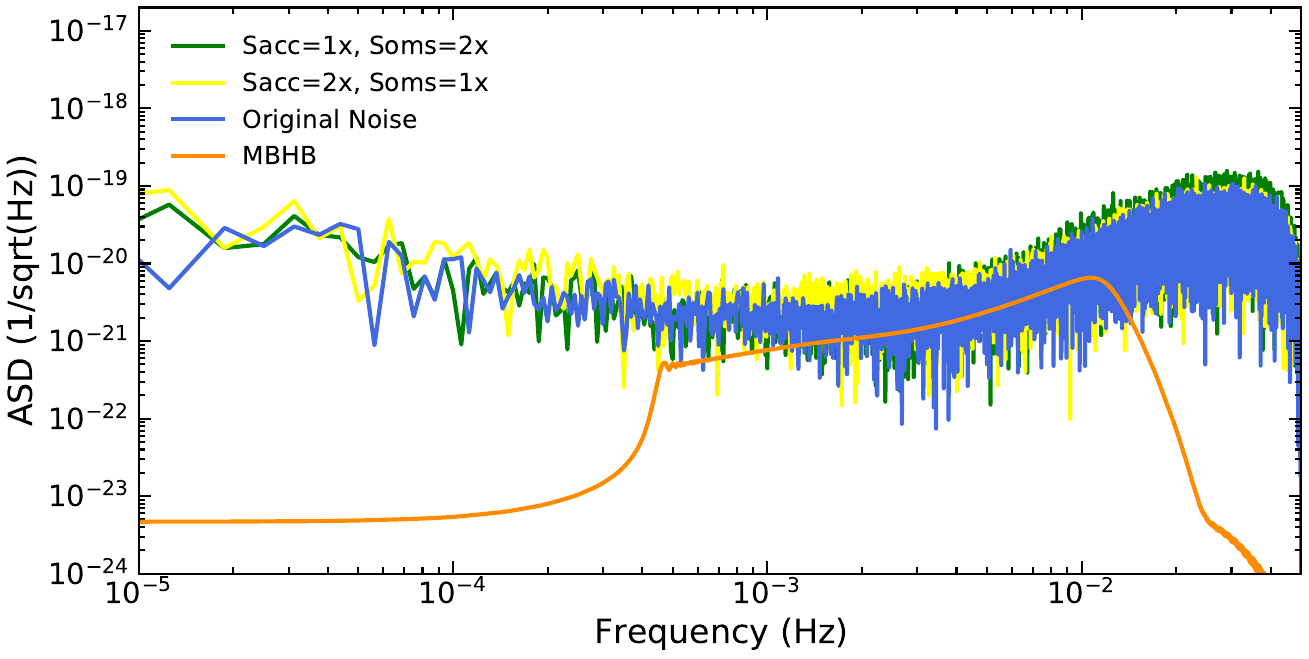}
\caption{\label{fig:noise3}The changes in the overall noise spectrum when the magnitudes of the optical metrology system noise and acceleration noise are altered. For our test data set, the optical noises affect the total SNR (SNR $\sim$ 35 in this example) more significantly. }
\end{figure}

While, given the predefined criterion, one finds that noises originating from the optical metrology system has a more significant impact on the final outcomes. 
Further investigation shows that this is because that, increasing the optical metrology system noise (concentrated in the relatively high frequency band) will significantly reduce the SNR of coalescing MBHB signals due to the specific forms of the PSDs of the total noise and expected signal, see Fig~\ref{fig:noise3} for illustrations. 
This, but not the variations in noise PSDs, will make it more difficult for the model to extract signals and produce the results shown Fig~\ref{fig:noise2}. and Tab ~\ref{tab:noise}.

\begin{table}[h]
\caption{\label{tab:noise}For time-varying noise auto-correlations, the ratio of overlaps $\geq$ 0.9 between the model-extracted signals and the template signals across the entire dataset. The last two columns represent the means and standard deviations of the overlap.}
\begin{ruledtabular}
\renewcommand{\arraystretch}{1.2}
\begin{tabular}{lcccc}
\( S_{\text{OMS}}(f) \) & \( S_{\text{ACC}}(f) \) & Overlaps $\geq$ 0.9 & Mean & Std. Dev. \\
\colrule
1x & 1x & 100.0\% & 0.997 & 0.005 \\
1x & 2x & 92.2\% & 0.962 & 0.117 \\
1x & 3x & 81.5\% & 0.923 & 0.158 \\
1x & 4x & 56.0\% & 0.842 & 0.205 \\
2x & 1x & 78.3\% & 0.894 & 0.224 \\
\end{tabular}
\end{ruledtabular}
\end{table}

\subsection{\label{sec:glitch}Glitches}

\begin{figure*}[htb]
\centering
\includegraphics[width=0.98\textwidth]{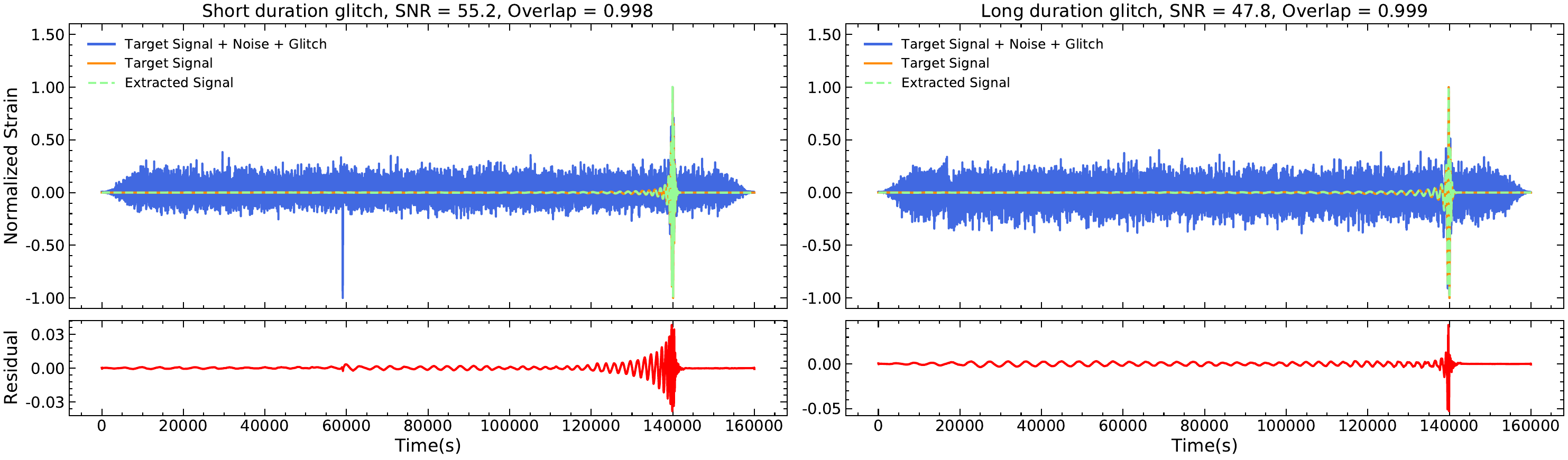}
\caption{\label{fig:glitch2}The left image displays the extraction result of our model when encountering a short duration glitch with a peak value of \(1 \times 10^{-13} \, \text{ms}^{-2}\). The right image shows the extraction result of our model when encountering a long duration glitch with a peak value of \(1 \times 10^{-14} \, \text{ms}^{-2}\).}
\end{figure*}

For instrumental transients, we consider the legacy model  from LISA Pathfinder for GRS glitches as representatives \cite{armano2022transient}
\begin{equation}
g(t) = \frac{\Delta v}{\tau_1 - \tau_2} \left( e^{\frac{-(t-t_0)}{\tau_1}} - e^{\frac{-(t-t_0)}{\tau_2}} \right) \Theta(t-t_0),
\label{eq:lpf}
\end{equation}
where \( t_0 \) is the time of glitch injection, \( \Delta v \) is the impulse transferred by the glitch (i.e., the gain of test mass velocity caused by the glitch), \( \tau_1, \tau_2 \) are time scales, and \(\Theta(t-t_0)\) denotes the Heaviside step function. 
Glitches are simulated and injected to {  the GRS residual acceleration noise $n_{\rm ACC}(t)$ of TM$_{12}$ (\emph{i.e.} the test-mass onboard spacecraft 1 and facing spacecraft 2). They are then passed through $T^{I}_{\rm ACC}(t)$ (in Eq. (\ref{eq:noise_transfer_function})), together with  the ``normal'' acceleration noises, to get the total TDI combinations with glitches.}
We considered two types of glitches according to Ref. \cite{armano2022transient}, that the short-duration glitches lasting approximately 70 seconds and long-duration glitches for about 3 hours. 
The initial amplitudes of both types of glitches were set to \(1 \times 10^{-14} \, \text{ms}^{-2}\),  see  Tab~\ref{tab:glitch_parm} for detailed parameters, and we tested the robustness of our model by increasing the amplitudes by 0.5 times for each new experiment until the performance fell below the threshold we set.
\begin{table}[h]
\caption{\label{tab:glitch_parm}The parameters corresponding to the two types of glitches, under this parameter setting, their amplitudes are both \(1 \times 10^{-14} \, \text{ms}^{-2}\).}
\begin{ruledtabular}
\renewcommand{\arraystretch}{1.2}
\begin{tabular}{lccc}
Duration & \( \Delta v \) (m/s) & \( \tau_1 \) (s) & \( \tau_2 \) (s) \\
\colrule
Short & \( 2.86 \times 10^{-13} \) & 10 & 11 \\
Long & \( 5.44 \times 10^{-11} \) & 2000 & 2001 \\
\end{tabular}
\end{ruledtabular}
\end{table}
  {All glitches were injected with an initial time $t_0=0$, and the time  difference between the injection points of the glitches and the mergers rang from 22 to 38 hours.}
Our model proficiently extracts the coalescing MBHB signals given such instrumental transients.
Fig~\ref{fig:glitch2} shows an example of the model's extraction result. 

\begin{figure*}[htb]
\centering
\includegraphics[width=0.98\textwidth]{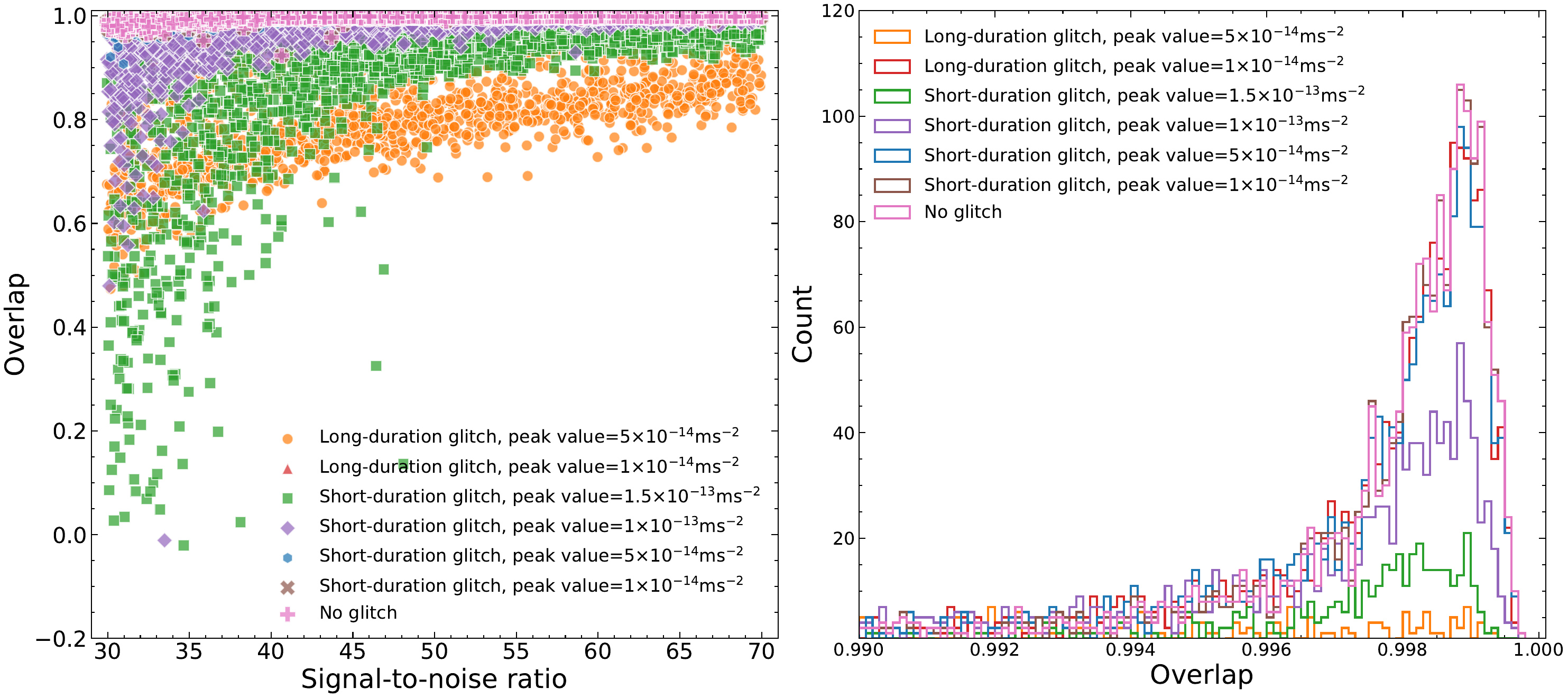}
\caption{\label{fig:glitch3}The left image displays the distributions of the overlaps for our test set confronting with glitches of varying intensities and duration. The image on the right presents the histogram for instances where the model achieves an overlap $\geq$ 0.99, corresponding to our test set confronting with glitches of varying intensities and duration.}
\end{figure*}
\begin{table}[h]
\caption{\label{tab:glitch}For glitches with different intensities and duration, the ratio of overlaps $\geq$ 0.9 between the model-extracted signals and the template signals across the entire dataset. The last two columns represent the means and standard deviations of the overlap.}
\begin{ruledtabular}
\renewcommand{\arraystretch}{1.2}
\begin{tabular}{lcccc}
Duration & Peak value(ms\(^{-2}\)) & Overlaps $\geq$ 0.9 & Mean & Std. Dev. \\
\colrule
None & None & 100.0\% & 0.997 & 0.005 \\
Short & \( 1 \times 10^{-14} \) & 100.0\% & 0.997 & 0.005\\
Short & \( 5 \times 10^{-14} \) & 99.9\% & 0.995 & 0.010\\
Short & \( 1 \times 10^{-13} \) & 94.9\% & 0.977 & 0.057\\
Short & \( 1.5 \times 10^{-13} \) & 72.0\% & 0.893 & 0.161\\
Long & \( 1 \times 10^{-14} \) & 100.0\% & 0.996 & 0.006\\
Long & \( 5 \times 10^{-14} \) & 31\% & 0.830 & 0.112\\
\end{tabular}
\end{ruledtabular}
\end{table}
The limits of our model's performance against glitches is shown in Fig~\ref{fig:glitch3}. 
The results indicate that the performance deteriorates as the duration of the glitch increases, and detailed results are presented in Tab~\ref{tab:glitch}. 
Specifically, for short-duration glitches, the maximum tolerable magnitude of the glitch is about \(1 \times 10^{-13} \, \text{ms}^{-2}\), while for long-duration glitches, the maximum tolerable magnitude is about \(1 \times 10^{-14} \, \text{ms}^{-2}\). We can see that the long-duration glitch has a greater impact on the extraction results. This is because we adjust its duration to be consistent with the duration of the merge phase of the MBHB signal, which greatly affects the model's judgment, resulting in erroneous extraction.
\subsection{\label{sec:mix}Mixing three types of non-stationarities}

\begin{table*}[htb]
\centering
\caption{\label{tab:mix}After mixing all anomalous or non-stationary conditions, the ratio of overlaps $\geq$ 0.9 between the model-extracted signals and the template signals across the entire dataset. The results also show the means and standard deviations of the obtained overlaps.}
\begin{tabularx}{\textwidth}{>{\centering\arraybackslash}X>{\centering\arraybackslash}X>{\centering\arraybackslash}X>{\centering\arraybackslash}X>{\centering\arraybackslash}X>{\centering\arraybackslash}X>{\centering\arraybackslash}X>{\centering\arraybackslash}X>{\centering\arraybackslash}X}
\toprule
\multicolumn{2}{c}{Data gaps} & \multicolumn{2}{c}{Non-stationary Gaussian noise} & \multicolumn{2}{c}{Glitches} & \multicolumn{3}{c}{Results} \\
\cmidrule(lr){1-2} \cmidrule(lr){3-4} \cmidrule(lr){5-6} \cmidrule(lr){7-9}
Scheduled gap & Unscheduled gap & \( S_{\text{OMS}}(f) \) & \( S_{\text{ACC}}(f) \) & Duration & Peak value (ms\(^{-2}\)) & Overlaps $\geq$ 0.9 & Mean & Std. Dev. \\
\midrule
$\checkmark$ & $\times$ & 1x & 3x & Short & \( 5 \times 10^{-14} \) & 81.6\% & 0.919 & 0.167 \\
$\checkmark$ & $\times$ & 1x & 3x & Long & \( 1 \times 10^{-14} \) & 80.5\% & 0.921 & 0.160 \\
$\times$ & Gap every 10 hours & 1x & 2x & Short & \( 5 \times 10^{-14} \) & 82.1\% & 0.914 & 0.194 \\
$\times$ & Gap every 10 hours & 1x & 2x & Long & \( 1 \times 10^{-14} \) & 81.8\% & 0.905 & 0.216 \\
\bottomrule
\end{tabularx}
\end{table*}

At last but not least, we conducted tests by mixing three types of non-stationarities and have listed in Tab~\ref{tab:mix} the results achieved by our model under different conditions. 
When all these combine, our model still exhibits considerable extraction capability. 
When long-term scheduled gap exists, our model can tolerate with time-varying noise PSDs together with short-duration glitches of maximum peak value \(\sim 5 \times 10^{-14} \, \text{ms}^{-2}\) or long-duration glitches  \(\sim 1 \times 10^{-14} \, \text{ms}^{-2}\).
When the occurrence frequency of unscheduled gaps reaches about once every 10 hours, our model can tolerate with relative weaker changes in noise (residual acceleration noise) PSDs combined with short-duration glitches of maximum peak value \(\sim 5 \times 10^{-14} \, \text{ms}^{-2}\) or long-duration glitches  \(\sim 1 \times 10^{-14} \, \text{ms}^{-2}\).
These tests give typically the possible extreme cases in real science operations and measurements of space-borne GW antennas, and our model have shown high robustness and feasibility in typical signal extractions. 
It can be observed that high-frequency unscheduled gaps may have important impacts, which is predominantly due to substantial loss of spectral information and further compounded by additional anomalies that significantly obfuscate the model's signal extraction capabilities. 
In future works, we will try to improve the model's ability to tolerate these mixed situations in order to improve performance in more extreme situations, and we will focus on unscheduled gaps, improving the model's tolerance for this situation and mitigating its impact on other anomalies.

\section{\label{sec:summary}Concluding remarks}

In this work, we developed a deep learning model for accurate GW extractions against possible data anomalies or non-stationarities for space-borne GW antennas, such as LISA and LISA-like missions, which also provide robustness studies of deep learning models in data processing of space-borne GW missions.
Our research focuses on the three types of non-stationarities, including gaps, glitches, and time varying noise auto-correlations.
Compared with the current state-of-the-art models, our model exhibits the same performance for the ideal and anomaly free cases in signal extractions.
While, confronted with the three types of non-stationarities and even their mixtures in some extreme cases, our model still shows considerable adaptability.
It is worth noting that when all anomalies occur at the same time, the model's tolerance for these anomalies would degrade compared with each stand-alone situation, and among these the unscheduled gap turns out to be a more important factor.


Deep learning has provided a method for the rapid detection of space-borne GW missions, and robustness research is key to ensuring that deep learning models are applicable in real detection scenarios. In the future, we will use this signal extraction model as a pre-processing model to reduce the difficulty of subsequent scientific analysis such as GW signal detection and parameter estimation through robust extraction of GW signals, becoming an important part of the Taiji detection pipeline. It is planned to process longer datasets in future works and to incorporate a greater variety of realistic simulation scenarios, including various levels of SNR and types of anomalies. 
Additionally, we will focus on unscheduled gaps, seeking new improvements to mitigate the substantial impact of such anomalies on model performance. 

\begin{acknowledgments}
This work is supported by the National Key Research and Development Program of China No. 2021YFC2201901, No. 2021YFC2201903, No. 2020YFC2200601 and No. 2020YFC2200901.
\end{acknowledgments}

\bibliography{apssamp}

\begin{thebibliography}{82}%
\makeatletter
\providecommand \@ifxundefined [1]{%
 \@ifx{#1\undefined}
}%
\providecommand \@ifnum [1]{%
 \ifnum #1\expandafter \@firstoftwo
 \else \expandafter \@secondoftwo
 \fi
}%
\providecommand \@ifx [1]{%
 \ifx #1\expandafter \@firstoftwo
 \else \expandafter \@secondoftwo
 \fi
}%
\providecommand \natexlab [1]{#1}%
\providecommand \enquote  [1]{``#1''}%
\providecommand \bibnamefont  [1]{#1}%
\providecommand \bibfnamefont [1]{#1}%
\providecommand \citenamefont [1]{#1}%
\providecommand \href@noop [0]{\@secondoftwo}%
\providecommand \href [0]{\begingroup \@sanitize@url \@href}%
\providecommand \@href[1]{\@@startlink{#1}\@@href}%
\providecommand \@@href[1]{\endgroup#1\@@endlink}%
\providecommand \@sanitize@url [0]{\catcode `\\12\catcode `\$12\catcode `\&12\catcode `\#12\catcode `\^12\catcode `\_12\catcode `\%12\relax}%
\providecommand \@@startlink[1]{}%
\providecommand \@@endlink[0]{}%
\providecommand \url  [0]{\begingroup\@sanitize@url \@url }%
\providecommand \@url [1]{\endgroup\@href {#1}{\urlprefix }}%
\providecommand \urlprefix  [0]{URL }%
\providecommand \Eprint [0]{\href }%
\providecommand \doibase [0]{https://doi.org/}%
\providecommand \selectlanguage [0]{\@gobble}%
\providecommand \bibinfo  [0]{\@secondoftwo}%
\providecommand \bibfield  [0]{\@secondoftwo}%
\providecommand \translation [1]{[#1]}%
\providecommand \BibitemOpen [0]{}%
\providecommand \bibitemStop [0]{}%
\providecommand \bibitemNoStop [0]{.\EOS\space}%
\providecommand \EOS [0]{\spacefactor3000\relax}%
\providecommand \BibitemShut  [1]{\csname bibitem#1\endcsname}%
\let\auto@bib@innerbib\@empty
\bibitem [{\citenamefont {Abbott}\ \emph {et~al.}(2016{\natexlab{a}})\citenamefont {Abbott}, \citenamefont {Abbott}, \citenamefont {Abbott}, \citenamefont {Abernathy}, \citenamefont {Acernese}, \citenamefont {Ackley}, \citenamefont {Adams}, \citenamefont {Adams}, \citenamefont {Addesso}, \citenamefont {Adhikari} \emph {et~al.}}]{abbott2016gw150914}%
  \BibitemOpen
  \bibfield  {author} {\bibinfo {author} {\bibfnamefont {B.~P.}\ \bibnamefont {Abbott}}, \bibinfo {author} {\bibfnamefont {R.}~\bibnamefont {Abbott}}, \bibinfo {author} {\bibfnamefont {T.}~\bibnamefont {Abbott}}, \bibinfo {author} {\bibfnamefont {M.}~\bibnamefont {Abernathy}}, \bibinfo {author} {\bibfnamefont {F.}~\bibnamefont {Acernese}}, \bibinfo {author} {\bibfnamefont {K.}~\bibnamefont {Ackley}}, \bibinfo {author} {\bibfnamefont {C.}~\bibnamefont {Adams}}, \bibinfo {author} {\bibfnamefont {T.}~\bibnamefont {Adams}}, \bibinfo {author} {\bibfnamefont {P.}~\bibnamefont {Addesso}}, \bibinfo {author} {\bibfnamefont {R.}~\bibnamefont {Adhikari}}, \emph {et~al.},\ }\bibfield  {title} {\bibinfo {title} {Gw150914: The advanced ligo detectors in the era of first discoveries},\ }\href@noop {} {\bibfield  {journal} {\bibinfo  {journal} {Physical review letters}\ }\textbf {\bibinfo {volume} {116}},\ \bibinfo {pages} {131103} (\bibinfo {year} {2016}{\natexlab{a}})}\BibitemShut {NoStop}%
\bibitem [{\citenamefont {Abbott}\ \emph {et~al.}(2016{\natexlab{b}})\citenamefont {Abbott}, \citenamefont {Abbott}, \citenamefont {Abbott}, \citenamefont {Abernathy}, \citenamefont {Acernese}, \citenamefont {Ackley}, \citenamefont {Adams}, \citenamefont {Adams}, \citenamefont {Addesso}, \citenamefont {Adhikari} \emph {et~al.}}]{abbott2016observation}%
  \BibitemOpen
  \bibfield  {author} {\bibinfo {author} {\bibfnamefont {B.~P.}\ \bibnamefont {Abbott}}, \bibinfo {author} {\bibfnamefont {R.}~\bibnamefont {Abbott}}, \bibinfo {author} {\bibfnamefont {T.}~\bibnamefont {Abbott}}, \bibinfo {author} {\bibfnamefont {M.}~\bibnamefont {Abernathy}}, \bibinfo {author} {\bibfnamefont {F.}~\bibnamefont {Acernese}}, \bibinfo {author} {\bibfnamefont {K.}~\bibnamefont {Ackley}}, \bibinfo {author} {\bibfnamefont {C.}~\bibnamefont {Adams}}, \bibinfo {author} {\bibfnamefont {T.}~\bibnamefont {Adams}}, \bibinfo {author} {\bibfnamefont {P.}~\bibnamefont {Addesso}}, \bibinfo {author} {\bibfnamefont {R.}~\bibnamefont {Adhikari}}, \emph {et~al.},\ }\bibfield  {title} {\bibinfo {title} {Observation of gravitational waves from a binary black hole merger},\ }\href@noop {} {\bibfield  {journal} {\bibinfo  {journal} {Physical review letters}\ }\textbf {\bibinfo {volume} {116}},\ \bibinfo {pages} {061102} (\bibinfo {year} {2016}{\natexlab{b}})}\BibitemShut {NoStop}%
\bibitem [{\citenamefont {Abbott}\ \emph {et~al.}(2016{\natexlab{c}})\citenamefont {Abbott}, \citenamefont {Abbott}, \citenamefont {Abbott}, \citenamefont {Abernathy}, \citenamefont {Acernese}, \citenamefont {Ackley}, \citenamefont {Adams}, \citenamefont {Adams}, \citenamefont {Addesso}, \citenamefont {Adhikari} \emph {et~al.}}]{abbott2016gw151226}%
  \BibitemOpen
  \bibfield  {author} {\bibinfo {author} {\bibfnamefont {B.~P.}\ \bibnamefont {Abbott}}, \bibinfo {author} {\bibfnamefont {R.}~\bibnamefont {Abbott}}, \bibinfo {author} {\bibfnamefont {T.}~\bibnamefont {Abbott}}, \bibinfo {author} {\bibfnamefont {M.}~\bibnamefont {Abernathy}}, \bibinfo {author} {\bibfnamefont {F.}~\bibnamefont {Acernese}}, \bibinfo {author} {\bibfnamefont {K.}~\bibnamefont {Ackley}}, \bibinfo {author} {\bibfnamefont {C.}~\bibnamefont {Adams}}, \bibinfo {author} {\bibfnamefont {T.}~\bibnamefont {Adams}}, \bibinfo {author} {\bibfnamefont {P.}~\bibnamefont {Addesso}}, \bibinfo {author} {\bibfnamefont {R.}~\bibnamefont {Adhikari}}, \emph {et~al.},\ }\bibfield  {title} {\bibinfo {title} {Gw151226: observation of gravitational waves from a 22-solar-mass binary black hole coalescence},\ }\href@noop {} {\bibfield  {journal} {\bibinfo  {journal} {Physical review letters}\ }\textbf {\bibinfo {volume} {116}},\ \bibinfo {pages} {241103} (\bibinfo {year} {2016}{\natexlab{c}})}\BibitemShut {NoStop}%
\bibitem [{\citenamefont {Abbott}\ \emph {et~al.}(2016{\natexlab{d}})\citenamefont {Abbott}, \citenamefont {Abbott}, \citenamefont {Abbott}, \citenamefont {Abernathy}, \citenamefont {Acernese}, \citenamefont {Ackley}, \citenamefont {Adams}, \citenamefont {Adams}, \citenamefont {Addesso}, \citenamefont {Adhikari} \emph {et~al.}}]{abbott2016binary}%
  \BibitemOpen
  \bibfield  {author} {\bibinfo {author} {\bibfnamefont {B.~P.}\ \bibnamefont {Abbott}}, \bibinfo {author} {\bibfnamefont {R.}~\bibnamefont {Abbott}}, \bibinfo {author} {\bibfnamefont {T.}~\bibnamefont {Abbott}}, \bibinfo {author} {\bibfnamefont {M.}~\bibnamefont {Abernathy}}, \bibinfo {author} {\bibfnamefont {F.}~\bibnamefont {Acernese}}, \bibinfo {author} {\bibfnamefont {K.}~\bibnamefont {Ackley}}, \bibinfo {author} {\bibfnamefont {C.}~\bibnamefont {Adams}}, \bibinfo {author} {\bibfnamefont {T.}~\bibnamefont {Adams}}, \bibinfo {author} {\bibfnamefont {P.}~\bibnamefont {Addesso}}, \bibinfo {author} {\bibfnamefont {R.}~\bibnamefont {Adhikari}}, \emph {et~al.},\ }\bibfield  {title} {\bibinfo {title} {Binary black hole mergers in the first advanced ligo observing run},\ }\href@noop {} {\bibfield  {journal} {\bibinfo  {journal} {Physical Review X}\ }\textbf {\bibinfo {volume} {6}},\ \bibinfo {pages} {041015} (\bibinfo {year} {2016}{\natexlab{d}})}\BibitemShut {NoStop}%
\bibitem [{\citenamefont {Scientific}\ \emph {et~al.}(2017)\citenamefont {Scientific}, \citenamefont {Abbott}, \citenamefont {Abbott}, \citenamefont {Abbott}, \citenamefont {Acernese}, \citenamefont {Ackley}, \citenamefont {Adams}, \citenamefont {Adams}, \citenamefont {Addesso}, \citenamefont {Adhikari} \emph {et~al.}}]{scientific2017gw170104}%
  \BibitemOpen
  \bibfield  {author} {\bibinfo {author} {\bibfnamefont {L.}~\bibnamefont {Scientific}}, \bibinfo {author} {\bibfnamefont {B.~P.}\ \bibnamefont {Abbott}}, \bibinfo {author} {\bibfnamefont {R.}~\bibnamefont {Abbott}}, \bibinfo {author} {\bibfnamefont {T.}~\bibnamefont {Abbott}}, \bibinfo {author} {\bibfnamefont {F.}~\bibnamefont {Acernese}}, \bibinfo {author} {\bibfnamefont {K.}~\bibnamefont {Ackley}}, \bibinfo {author} {\bibfnamefont {C.}~\bibnamefont {Adams}}, \bibinfo {author} {\bibfnamefont {T.}~\bibnamefont {Adams}}, \bibinfo {author} {\bibfnamefont {P.}~\bibnamefont {Addesso}}, \bibinfo {author} {\bibfnamefont {R.}~\bibnamefont {Adhikari}}, \emph {et~al.},\ }\bibfield  {title} {\bibinfo {title} {Gw170104: observation of a 50-solar-mass binary black hole coalescence at redshift 0.2},\ }\href@noop {} {\bibfield  {journal} {\bibinfo  {journal} {Physical review letters}\ }\textbf {\bibinfo {volume} {118}},\ \bibinfo {pages} {221101} (\bibinfo {year} {2017})}\BibitemShut {NoStop}%
\bibitem [{\citenamefont {Abbott}\ \emph {et~al.}(2017{\natexlab{a}})\citenamefont {Abbott}, \citenamefont {Abbott}, \citenamefont {Abbott}, \citenamefont {Acernese}, \citenamefont {Ackley}, \citenamefont {Adams}, \citenamefont {Adams}, \citenamefont {Addesso}, \citenamefont {Adhikari}, \citenamefont {Adya} \emph {et~al.}}]{abbott2017gw170608}%
  \BibitemOpen
  \bibfield  {author} {\bibinfo {author} {\bibfnamefont {B.~P.}\ \bibnamefont {Abbott}}, \bibinfo {author} {\bibfnamefont {R.}~\bibnamefont {Abbott}}, \bibinfo {author} {\bibfnamefont {T.}~\bibnamefont {Abbott}}, \bibinfo {author} {\bibfnamefont {F.}~\bibnamefont {Acernese}}, \bibinfo {author} {\bibfnamefont {K.}~\bibnamefont {Ackley}}, \bibinfo {author} {\bibfnamefont {C.}~\bibnamefont {Adams}}, \bibinfo {author} {\bibfnamefont {T.}~\bibnamefont {Adams}}, \bibinfo {author} {\bibfnamefont {P.}~\bibnamefont {Addesso}}, \bibinfo {author} {\bibfnamefont {R.}~\bibnamefont {Adhikari}}, \bibinfo {author} {\bibfnamefont {V.}~\bibnamefont {Adya}}, \emph {et~al.},\ }\bibfield  {title} {\bibinfo {title} {Gw170608: observation of a 19 solar-mass binary black hole coalescence},\ }\href@noop {} {\bibfield  {journal} {\bibinfo  {journal} {The Astrophysical Journal Letters}\ }\textbf {\bibinfo {volume} {851}},\ \bibinfo {pages} {L35} (\bibinfo {year} {2017}{\natexlab{a}})}\BibitemShut {NoStop}%
\bibitem [{\citenamefont {Abbott}\ \emph {et~al.}(2017{\natexlab{b}})\citenamefont {Abbott}, \citenamefont {Abbott}, \citenamefont {Abbott}, \citenamefont {Acernese}, \citenamefont {Ackley}, \citenamefont {Adams}, \citenamefont {Adams}, \citenamefont {Addesso}, \citenamefont {Adhikari}, \citenamefont {Adya} \emph {et~al.}}]{abbott2017gw170814}%
  \BibitemOpen
  \bibfield  {author} {\bibinfo {author} {\bibfnamefont {B.~P.}\ \bibnamefont {Abbott}}, \bibinfo {author} {\bibfnamefont {R.}~\bibnamefont {Abbott}}, \bibinfo {author} {\bibfnamefont {T.}~\bibnamefont {Abbott}}, \bibinfo {author} {\bibfnamefont {F.}~\bibnamefont {Acernese}}, \bibinfo {author} {\bibfnamefont {K.}~\bibnamefont {Ackley}}, \bibinfo {author} {\bibfnamefont {C.}~\bibnamefont {Adams}}, \bibinfo {author} {\bibfnamefont {T.}~\bibnamefont {Adams}}, \bibinfo {author} {\bibfnamefont {P.}~\bibnamefont {Addesso}}, \bibinfo {author} {\bibfnamefont {R.~X.}\ \bibnamefont {Adhikari}}, \bibinfo {author} {\bibfnamefont {V.~B.}\ \bibnamefont {Adya}}, \emph {et~al.},\ }\bibfield  {title} {\bibinfo {title} {Gw170814: a three-detector observation of gravitational waves from a binary black hole coalescence},\ }\href@noop {} {\bibfield  {journal} {\bibinfo  {journal} {Physical review letters}\ }\textbf {\bibinfo {volume} {119}},\ \bibinfo {pages} {141101} (\bibinfo {year} {2017}{\natexlab{b}})}\BibitemShut {NoStop}%
\bibitem [{\citenamefont {Abbott}\ \emph {et~al.}(2017{\natexlab{c}})\citenamefont {Abbott}, \citenamefont {Abbott}, \citenamefont {Abbott}, \citenamefont {Acernese}, \citenamefont {Ackley}, \citenamefont {Adams}, \citenamefont {Adams}, \citenamefont {Addesso}, \citenamefont {Adhikari}, \citenamefont {Adya} \emph {et~al.}}]{abbott2017gw170817}%
  \BibitemOpen
  \bibfield  {author} {\bibinfo {author} {\bibfnamefont {B.~P.}\ \bibnamefont {Abbott}}, \bibinfo {author} {\bibfnamefont {R.}~\bibnamefont {Abbott}}, \bibinfo {author} {\bibfnamefont {T.}~\bibnamefont {Abbott}}, \bibinfo {author} {\bibfnamefont {F.}~\bibnamefont {Acernese}}, \bibinfo {author} {\bibfnamefont {K.}~\bibnamefont {Ackley}}, \bibinfo {author} {\bibfnamefont {C.}~\bibnamefont {Adams}}, \bibinfo {author} {\bibfnamefont {T.}~\bibnamefont {Adams}}, \bibinfo {author} {\bibfnamefont {P.}~\bibnamefont {Addesso}}, \bibinfo {author} {\bibfnamefont {R.}~\bibnamefont {Adhikari}}, \bibinfo {author} {\bibfnamefont {V.~B.}\ \bibnamefont {Adya}}, \emph {et~al.},\ }\bibfield  {title} {\bibinfo {title} {Gw170817: observation of gravitational waves from a binary neutron star inspiral},\ }\href@noop {} {\bibfield  {journal} {\bibinfo  {journal} {Physical review letters}\ }\textbf {\bibinfo {volume} {119}},\ \bibinfo {pages} {161101} (\bibinfo {year} {2017}{\natexlab{c}})}\BibitemShut {NoStop}%
\bibitem [{\citenamefont {Abbott}\ \emph {et~al.}(2020{\natexlab{a}})\citenamefont {Abbott}, \citenamefont {Abbott}, \citenamefont {Abbott}, \citenamefont {Abraham}, \citenamefont {Acernese}, \citenamefont {Ackley}, \citenamefont {Adams}, \citenamefont {Adhikari}, \citenamefont {Adya}, \citenamefont {Affeldt} \emph {et~al.}}]{abbott2020gw190425}%
  \BibitemOpen
  \bibfield  {author} {\bibinfo {author} {\bibfnamefont {B.}~\bibnamefont {Abbott}}, \bibinfo {author} {\bibfnamefont {R.}~\bibnamefont {Abbott}}, \bibinfo {author} {\bibfnamefont {T.}~\bibnamefont {Abbott}}, \bibinfo {author} {\bibfnamefont {S.}~\bibnamefont {Abraham}}, \bibinfo {author} {\bibfnamefont {F.}~\bibnamefont {Acernese}}, \bibinfo {author} {\bibfnamefont {K.}~\bibnamefont {Ackley}}, \bibinfo {author} {\bibfnamefont {C.}~\bibnamefont {Adams}}, \bibinfo {author} {\bibfnamefont {R.}~\bibnamefont {Adhikari}}, \bibinfo {author} {\bibfnamefont {V.}~\bibnamefont {Adya}}, \bibinfo {author} {\bibfnamefont {C.}~\bibnamefont {Affeldt}}, \emph {et~al.},\ }\bibfield  {title} {\bibinfo {title} {Gw190425: Observation of a compact binary coalescence with total mass~ 3.4 m},\ }\href@noop {} {\bibfield  {journal} {\bibinfo  {journal} {The Astrophysical Journal}\ }\textbf {\bibinfo {volume} {892}},\ \bibinfo {pages} {L3} (\bibinfo {year} {2020}{\natexlab{a}})}\BibitemShut {NoStop}%
\bibitem [{\citenamefont {Abbott}\ \emph {et~al.}(2020{\natexlab{b}})\citenamefont {Abbott}, \citenamefont {Abbott}, \citenamefont {Abraham}, \citenamefont {Acernese}, \citenamefont {Ackley}, \citenamefont {Adams}, \citenamefont {Adhikari}, \citenamefont {Adya}, \citenamefont {Affeldt}, \citenamefont {Agathos} \emph {et~al.}}]{abbott2020gw190412}%
  \BibitemOpen
  \bibfield  {author} {\bibinfo {author} {\bibfnamefont {R.}~\bibnamefont {Abbott}}, \bibinfo {author} {\bibfnamefont {T.}~\bibnamefont {Abbott}}, \bibinfo {author} {\bibfnamefont {S.}~\bibnamefont {Abraham}}, \bibinfo {author} {\bibfnamefont {F.}~\bibnamefont {Acernese}}, \bibinfo {author} {\bibfnamefont {K.}~\bibnamefont {Ackley}}, \bibinfo {author} {\bibfnamefont {C.}~\bibnamefont {Adams}}, \bibinfo {author} {\bibfnamefont {R.~X.}\ \bibnamefont {Adhikari}}, \bibinfo {author} {\bibfnamefont {V.}~\bibnamefont {Adya}}, \bibinfo {author} {\bibfnamefont {C.}~\bibnamefont {Affeldt}}, \bibinfo {author} {\bibfnamefont {M.}~\bibnamefont {Agathos}}, \emph {et~al.},\ }\bibfield  {title} {\bibinfo {title} {Gw190412: Observation of a binary-black-hole coalescence with asymmetric masses},\ }\href@noop {} {\bibfield  {journal} {\bibinfo  {journal} {Physical Review D}\ }\textbf {\bibinfo {volume} {102}},\ \bibinfo {pages} {043015} (\bibinfo {year} {2020}{\natexlab{b}})}\BibitemShut {NoStop}%
\bibitem [{\citenamefont {Abbott}\ \emph {et~al.}(2020{\natexlab{c}})\citenamefont {Abbott}, \citenamefont {Abbott}, \citenamefont {Abraham}, \citenamefont {Acernese}, \citenamefont {Ackley}, \citenamefont {Adams}, \citenamefont {Adhikari}, \citenamefont {Adya}, \citenamefont {Affeldt}, \citenamefont {Agathos} \emph {et~al.}}]{abbott2020gw190521}%
  \BibitemOpen
  \bibfield  {author} {\bibinfo {author} {\bibfnamefont {R.}~\bibnamefont {Abbott}}, \bibinfo {author} {\bibfnamefont {T.}~\bibnamefont {Abbott}}, \bibinfo {author} {\bibfnamefont {S.}~\bibnamefont {Abraham}}, \bibinfo {author} {\bibfnamefont {F.}~\bibnamefont {Acernese}}, \bibinfo {author} {\bibfnamefont {K.}~\bibnamefont {Ackley}}, \bibinfo {author} {\bibfnamefont {C.}~\bibnamefont {Adams}}, \bibinfo {author} {\bibfnamefont {R.}~\bibnamefont {Adhikari}}, \bibinfo {author} {\bibfnamefont {V.}~\bibnamefont {Adya}}, \bibinfo {author} {\bibfnamefont {C.}~\bibnamefont {Affeldt}}, \bibinfo {author} {\bibfnamefont {M.}~\bibnamefont {Agathos}}, \emph {et~al.},\ }\bibfield  {title} {\bibinfo {title} {Gw190521: a binary black hole merger with a total mass of 150 m},\ }\href@noop {} {\bibfield  {journal} {\bibinfo  {journal} {Physical review letters}\ }\textbf {\bibinfo {volume} {125}},\ \bibinfo {pages} {101102} (\bibinfo {year} {2020}{\natexlab{c}})}\BibitemShut {NoStop}%
\bibitem [{\citenamefont {Abbott}\ \emph {et~al.}(2021)\citenamefont {Abbott}, \citenamefont {Abbott}, \citenamefont {Abraham}, \citenamefont {Acernese}, \citenamefont {Ackley}, \citenamefont {Adams}, \citenamefont {Adams}, \citenamefont {Adhikari}, \citenamefont {Adya}, \citenamefont {Affeldt} \emph {et~al.}}]{abbott2021tests}%
  \BibitemOpen
  \bibfield  {author} {\bibinfo {author} {\bibfnamefont {R.}~\bibnamefont {Abbott}}, \bibinfo {author} {\bibfnamefont {T.}~\bibnamefont {Abbott}}, \bibinfo {author} {\bibfnamefont {S.}~\bibnamefont {Abraham}}, \bibinfo {author} {\bibfnamefont {F.}~\bibnamefont {Acernese}}, \bibinfo {author} {\bibfnamefont {K.}~\bibnamefont {Ackley}}, \bibinfo {author} {\bibfnamefont {A.}~\bibnamefont {Adams}}, \bibinfo {author} {\bibfnamefont {C.}~\bibnamefont {Adams}}, \bibinfo {author} {\bibfnamefont {R.~X.}\ \bibnamefont {Adhikari}}, \bibinfo {author} {\bibfnamefont {V.}~\bibnamefont {Adya}}, \bibinfo {author} {\bibfnamefont {C.}~\bibnamefont {Affeldt}}, \emph {et~al.},\ }\bibfield  {title} {\bibinfo {title} {Tests of general relativity with binary black holes from the second ligo-virgo gravitational-wave transient catalog},\ }\href@noop {} {\bibfield  {journal} {\bibinfo  {journal} {Physical review D}\ }\textbf {\bibinfo {volume} {103}},\ \bibinfo {pages} {122002} (\bibinfo {year} {2021})}\BibitemShut {NoStop}%
\bibitem [{\citenamefont {Ezquiaga}(2021)}]{ezquiaga2021hearing}%
  \BibitemOpen
  \bibfield  {author} {\bibinfo {author} {\bibfnamefont {J.~M.}\ \bibnamefont {Ezquiaga}},\ }\bibfield  {title} {\bibinfo {title} {Hearing gravity from the cosmos: Gwtc-2 probes general relativity at cosmological scales},\ }\href@noop {} {\bibfield  {journal} {\bibinfo  {journal} {Physics Letters B}\ }\textbf {\bibinfo {volume} {822}},\ \bibinfo {pages} {136665} (\bibinfo {year} {2021})}\BibitemShut {NoStop}%
\bibitem [{\citenamefont {Bozzola}\ and\ \citenamefont {Paschalidis}(2021)}]{bozzola2021general}%
  \BibitemOpen
  \bibfield  {author} {\bibinfo {author} {\bibfnamefont {G.}~\bibnamefont {Bozzola}}\ and\ \bibinfo {author} {\bibfnamefont {V.}~\bibnamefont {Paschalidis}},\ }\bibfield  {title} {\bibinfo {title} {General relativistic simulations of the quasicircular inspiral and merger of charged black holes: Gw150914 and fundamental physics implications},\ }\href@noop {} {\bibfield  {journal} {\bibinfo  {journal} {Physical Review Letters}\ }\textbf {\bibinfo {volume} {126}},\ \bibinfo {pages} {041103} (\bibinfo {year} {2021})}\BibitemShut {NoStop}%
\bibitem [{\citenamefont {Abbott}\ \emph {et~al.}(2019)\citenamefont {Abbott}, \citenamefont {Abbott}, \citenamefont {Abbott}, \citenamefont {Abraham}, \citenamefont {Acernese}, \citenamefont {Ackley}, \citenamefont {Adams}, \citenamefont {Adhikari}, \citenamefont {Adya}, \citenamefont {Affeldt} \emph {et~al.}}]{abbott2019gwtc}%
  \BibitemOpen
  \bibfield  {author} {\bibinfo {author} {\bibfnamefont {B.}~\bibnamefont {Abbott}}, \bibinfo {author} {\bibfnamefont {R.}~\bibnamefont {Abbott}}, \bibinfo {author} {\bibfnamefont {T.}~\bibnamefont {Abbott}}, \bibinfo {author} {\bibfnamefont {S.}~\bibnamefont {Abraham}}, \bibinfo {author} {\bibfnamefont {F.}~\bibnamefont {Acernese}}, \bibinfo {author} {\bibfnamefont {K.}~\bibnamefont {Ackley}}, \bibinfo {author} {\bibfnamefont {C.}~\bibnamefont {Adams}}, \bibinfo {author} {\bibfnamefont {R.}~\bibnamefont {Adhikari}}, \bibinfo {author} {\bibfnamefont {V.}~\bibnamefont {Adya}}, \bibinfo {author} {\bibfnamefont {C.}~\bibnamefont {Affeldt}}, \emph {et~al.},\ }\bibfield  {title} {\bibinfo {title} {Gwtc-1: a gravitational-wave transient catalog of compact binary mergers observed by ligo and virgo during the first and second observing runs},\ }\href@noop {} {\bibfield  {journal} {\bibinfo  {journal} {Physical Review X}\ }\textbf {\bibinfo {volume} {9}},\ \bibinfo {pages} {031040} (\bibinfo {year}
  {2019})}\BibitemShut {NoStop}%
\bibitem [{\citenamefont {Abbott}\ \emph {et~al.}(2020{\natexlab{d}})\citenamefont {Abbott}, \citenamefont {Abbott}, \citenamefont {Abbott}, \citenamefont {Abraham}, \citenamefont {Acernese}, \citenamefont {Ackley}, \citenamefont {Adams}, \citenamefont {Adya}, \citenamefont {Affeldt}, \citenamefont {Agathos} \emph {et~al.}}]{abbott2020prospects}%
  \BibitemOpen
  \bibfield  {author} {\bibinfo {author} {\bibfnamefont {B.~P.}\ \bibnamefont {Abbott}}, \bibinfo {author} {\bibfnamefont {R.}~\bibnamefont {Abbott}}, \bibinfo {author} {\bibfnamefont {T.}~\bibnamefont {Abbott}}, \bibinfo {author} {\bibfnamefont {S.}~\bibnamefont {Abraham}}, \bibinfo {author} {\bibfnamefont {F.}~\bibnamefont {Acernese}}, \bibinfo {author} {\bibfnamefont {K.}~\bibnamefont {Ackley}}, \bibinfo {author} {\bibfnamefont {C.}~\bibnamefont {Adams}}, \bibinfo {author} {\bibfnamefont {V.}~\bibnamefont {Adya}}, \bibinfo {author} {\bibfnamefont {C.}~\bibnamefont {Affeldt}}, \bibinfo {author} {\bibfnamefont {M.}~\bibnamefont {Agathos}}, \emph {et~al.},\ }\bibfield  {title} {\bibinfo {title} {Prospects for observing and localizing gravitational-wave transients with advanced ligo, advanced virgo and kagra},\ }\href@noop {} {\bibfield  {journal} {\bibinfo  {journal} {Living reviews in relativity}\ }\textbf {\bibinfo {volume} {23}},\ \bibinfo {pages} {1} (\bibinfo {year} {2020}{\natexlab{d}})}\BibitemShut
  {NoStop}%
\bibitem [{\citenamefont {Freise}\ and\ \citenamefont {Strain}(2010)}]{freise2010interferometer}%
  \BibitemOpen
  \bibfield  {author} {\bibinfo {author} {\bibfnamefont {A.}~\bibnamefont {Freise}}\ and\ \bibinfo {author} {\bibfnamefont {K.}~\bibnamefont {Strain}},\ }\bibfield  {title} {\bibinfo {title} {Interferometer techniques for gravitational-wave detection},\ }\href@noop {} {\bibfield  {journal} {\bibinfo  {journal} {Living Reviews in Relativity}\ }\textbf {\bibinfo {volume} {13}},\ \bibinfo {pages} {1} (\bibinfo {year} {2010})}\BibitemShut {NoStop}%
\bibitem [{\citenamefont {Amaro-Seoane}\ \emph {et~al.}(2017)\citenamefont {Amaro-Seoane}, \citenamefont {Audley}, \citenamefont {Babak}, \citenamefont {Baker}, \citenamefont {Barausse}, \citenamefont {Bender}, \citenamefont {Berti}, \citenamefont {Binetruy}, \citenamefont {Born}, \citenamefont {Bortoluzzi} \emph {et~al.}}]{amaro2017laser}%
  \BibitemOpen
  \bibfield  {author} {\bibinfo {author} {\bibfnamefont {P.}~\bibnamefont {Amaro-Seoane}}, \bibinfo {author} {\bibfnamefont {H.}~\bibnamefont {Audley}}, \bibinfo {author} {\bibfnamefont {S.}~\bibnamefont {Babak}}, \bibinfo {author} {\bibfnamefont {J.}~\bibnamefont {Baker}}, \bibinfo {author} {\bibfnamefont {E.}~\bibnamefont {Barausse}}, \bibinfo {author} {\bibfnamefont {P.}~\bibnamefont {Bender}}, \bibinfo {author} {\bibfnamefont {E.}~\bibnamefont {Berti}}, \bibinfo {author} {\bibfnamefont {P.}~\bibnamefont {Binetruy}}, \bibinfo {author} {\bibfnamefont {M.}~\bibnamefont {Born}}, \bibinfo {author} {\bibfnamefont {D.}~\bibnamefont {Bortoluzzi}}, \emph {et~al.},\ }\bibfield  {title} {\bibinfo {title} {Laser interferometer space antenna},\ }\href@noop {} {\bibfield  {journal} {\bibinfo  {journal} {arXiv preprint arXiv:1702.00786}\ } (\bibinfo {year} {2017})}\BibitemShut {NoStop}%
\bibitem [{\citenamefont {Baker}\ \emph {et~al.}(2019)\citenamefont {Baker}, \citenamefont {Bellovary}, \citenamefont {Bender}, \citenamefont {Berti}, \citenamefont {Caldwell}, \citenamefont {Camp}, \citenamefont {Conklin}, \citenamefont {Cornish}, \citenamefont {Cutler}, \citenamefont {DeRosa} \emph {et~al.}}]{baker2019laser}%
  \BibitemOpen
  \bibfield  {author} {\bibinfo {author} {\bibfnamefont {J.}~\bibnamefont {Baker}}, \bibinfo {author} {\bibfnamefont {J.}~\bibnamefont {Bellovary}}, \bibinfo {author} {\bibfnamefont {P.~L.}\ \bibnamefont {Bender}}, \bibinfo {author} {\bibfnamefont {E.}~\bibnamefont {Berti}}, \bibinfo {author} {\bibfnamefont {R.}~\bibnamefont {Caldwell}}, \bibinfo {author} {\bibfnamefont {J.}~\bibnamefont {Camp}}, \bibinfo {author} {\bibfnamefont {J.~W.}\ \bibnamefont {Conklin}}, \bibinfo {author} {\bibfnamefont {N.}~\bibnamefont {Cornish}}, \bibinfo {author} {\bibfnamefont {C.}~\bibnamefont {Cutler}}, \bibinfo {author} {\bibfnamefont {R.}~\bibnamefont {DeRosa}}, \emph {et~al.},\ }\bibfield  {title} {\bibinfo {title} {The laser interferometer space antenna: unveiling the millihertz gravitational wave sky},\ }\href@noop {} {\bibfield  {journal} {\bibinfo  {journal} {arXiv preprint arXiv:1907.06482}\ } (\bibinfo {year} {2019})}\BibitemShut {NoStop}%
\bibitem [{\citenamefont {Gong}\ \emph {et~al.}(2011)\citenamefont {Gong}, \citenamefont {Xu}, \citenamefont {Bai}, \citenamefont {Cao}, \citenamefont {Chen}, \citenamefont {Chen}, \citenamefont {He}, \citenamefont {Heinzel}, \citenamefont {Lau}, \citenamefont {Liu} \emph {et~al.}}]{gong2011scientific}%
  \BibitemOpen
  \bibfield  {author} {\bibinfo {author} {\bibfnamefont {X.}~\bibnamefont {Gong}}, \bibinfo {author} {\bibfnamefont {S.}~\bibnamefont {Xu}}, \bibinfo {author} {\bibfnamefont {S.}~\bibnamefont {Bai}}, \bibinfo {author} {\bibfnamefont {Z.}~\bibnamefont {Cao}}, \bibinfo {author} {\bibfnamefont {G.}~\bibnamefont {Chen}}, \bibinfo {author} {\bibfnamefont {Y.}~\bibnamefont {Chen}}, \bibinfo {author} {\bibfnamefont {X.}~\bibnamefont {He}}, \bibinfo {author} {\bibfnamefont {G.}~\bibnamefont {Heinzel}}, \bibinfo {author} {\bibfnamefont {Y.-K.}\ \bibnamefont {Lau}}, \bibinfo {author} {\bibfnamefont {C.}~\bibnamefont {Liu}}, \emph {et~al.},\ }\bibfield  {title} {\bibinfo {title} {A scientific case study of an advanced lisa mission},\ }\href@noop {} {\bibfield  {journal} {\bibinfo  {journal} {Classical and Quantum Gravity}\ }\textbf {\bibinfo {volume} {28}},\ \bibinfo {pages} {094012} (\bibinfo {year} {2011})}\BibitemShut {NoStop}%
\bibitem [{\citenamefont {Hu}\ and\ \citenamefont {Wu}(2017)}]{hu2017taiji}%
  \BibitemOpen
  \bibfield  {author} {\bibinfo {author} {\bibfnamefont {W.-R.}\ \bibnamefont {Hu}}\ and\ \bibinfo {author} {\bibfnamefont {Y.-L.}\ \bibnamefont {Wu}},\ }\href@noop {} {\bibinfo {title} {The taiji program in space for gravitational wave physics and the nature of gravity}} (\bibinfo {year} {2017})\BibitemShut {NoStop}%
\bibitem [{\citenamefont {Luo}\ \emph {et~al.}(2016)\citenamefont {Luo}, \citenamefont {Chen}, \citenamefont {Duan}, \citenamefont {Gong}, \citenamefont {Hu}, \citenamefont {Ji}, \citenamefont {Liu}, \citenamefont {Mei}, \citenamefont {Milyukov}, \citenamefont {Sazhin} \emph {et~al.}}]{luo2016tianqin}%
  \BibitemOpen
  \bibfield  {author} {\bibinfo {author} {\bibfnamefont {J.}~\bibnamefont {Luo}}, \bibinfo {author} {\bibfnamefont {L.-S.}\ \bibnamefont {Chen}}, \bibinfo {author} {\bibfnamefont {H.-Z.}\ \bibnamefont {Duan}}, \bibinfo {author} {\bibfnamefont {Y.-G.}\ \bibnamefont {Gong}}, \bibinfo {author} {\bibfnamefont {S.}~\bibnamefont {Hu}}, \bibinfo {author} {\bibfnamefont {J.}~\bibnamefont {Ji}}, \bibinfo {author} {\bibfnamefont {Q.}~\bibnamefont {Liu}}, \bibinfo {author} {\bibfnamefont {J.}~\bibnamefont {Mei}}, \bibinfo {author} {\bibfnamefont {V.}~\bibnamefont {Milyukov}}, \bibinfo {author} {\bibfnamefont {M.}~\bibnamefont {Sazhin}}, \emph {et~al.},\ }\bibfield  {title} {\bibinfo {title} {Tianqin: a space-borne gravitational wave detector},\ }\href@noop {} {\bibfield  {journal} {\bibinfo  {journal} {Classical and Quantum Gravity}\ }\textbf {\bibinfo {volume} {33}},\ \bibinfo {pages} {035010} (\bibinfo {year} {2016})}\BibitemShut {NoStop}%
\bibitem [{\citenamefont {Dey}\ \emph {et~al.}(2021)\citenamefont {Dey}, \citenamefont {Karnesis}, \citenamefont {Toubiana}, \citenamefont {Barausse}, \citenamefont {Korsakova}, \citenamefont {Baghi},\ and\ \citenamefont {Basak}}]{dey2021effect}%
  \BibitemOpen
  \bibfield  {author} {\bibinfo {author} {\bibfnamefont {K.}~\bibnamefont {Dey}}, \bibinfo {author} {\bibfnamefont {N.}~\bibnamefont {Karnesis}}, \bibinfo {author} {\bibfnamefont {A.}~\bibnamefont {Toubiana}}, \bibinfo {author} {\bibfnamefont {E.}~\bibnamefont {Barausse}}, \bibinfo {author} {\bibfnamefont {N.}~\bibnamefont {Korsakova}}, \bibinfo {author} {\bibfnamefont {Q.}~\bibnamefont {Baghi}},\ and\ \bibinfo {author} {\bibfnamefont {S.}~\bibnamefont {Basak}},\ }\bibfield  {title} {\bibinfo {title} {Effect of data gaps on the detectability and parameter estimation of massive black hole binaries with lisa},\ }\href@noop {} {\bibfield  {journal} {\bibinfo  {journal} {Physical Review D}\ }\textbf {\bibinfo {volume} {104}},\ \bibinfo {pages} {044035} (\bibinfo {year} {2021})}\BibitemShut {NoStop}%
\bibitem [{\citenamefont {Baghi}\ \emph {et~al.}(2022)\citenamefont {Baghi}, \citenamefont {Korsakova}, \citenamefont {Slutsky}, \citenamefont {Castelli}, \citenamefont {Karnesis},\ and\ \citenamefont {Bayle}}]{baghi2022detection}%
  \BibitemOpen
  \bibfield  {author} {\bibinfo {author} {\bibfnamefont {Q.}~\bibnamefont {Baghi}}, \bibinfo {author} {\bibfnamefont {N.}~\bibnamefont {Korsakova}}, \bibinfo {author} {\bibfnamefont {J.}~\bibnamefont {Slutsky}}, \bibinfo {author} {\bibfnamefont {E.}~\bibnamefont {Castelli}}, \bibinfo {author} {\bibfnamefont {N.}~\bibnamefont {Karnesis}},\ and\ \bibinfo {author} {\bibfnamefont {J.-B.}\ \bibnamefont {Bayle}},\ }\bibfield  {title} {\bibinfo {title} {Detection and characterization of instrumental transients in lisa pathfinder and their projection to lisa},\ }\href@noop {} {\bibfield  {journal} {\bibinfo  {journal} {Physical Review D}\ }\textbf {\bibinfo {volume} {105}},\ \bibinfo {pages} {042002} (\bibinfo {year} {2022})}\BibitemShut {NoStop}%
\bibitem [{\citenamefont {Edwards}\ \emph {et~al.}(2020)\citenamefont {Edwards}, \citenamefont {Maturana-Russel}, \citenamefont {Meyer}, \citenamefont {Gair}, \citenamefont {Korsakova},\ and\ \citenamefont {Christensen}}]{edwards2020identifying}%
  \BibitemOpen
  \bibfield  {author} {\bibinfo {author} {\bibfnamefont {M.~C.}\ \bibnamefont {Edwards}}, \bibinfo {author} {\bibfnamefont {P.}~\bibnamefont {Maturana-Russel}}, \bibinfo {author} {\bibfnamefont {R.}~\bibnamefont {Meyer}}, \bibinfo {author} {\bibfnamefont {J.}~\bibnamefont {Gair}}, \bibinfo {author} {\bibfnamefont {N.}~\bibnamefont {Korsakova}},\ and\ \bibinfo {author} {\bibfnamefont {N.}~\bibnamefont {Christensen}},\ }\bibfield  {title} {\bibinfo {title} {Identifying and addressing nonstationary lisa noise},\ }\href@noop {} {\bibfield  {journal} {\bibinfo  {journal} {Physical Review D}\ }\textbf {\bibinfo {volume} {102}},\ \bibinfo {pages} {084062} (\bibinfo {year} {2020})}\BibitemShut {NoStop}%
\bibitem [{\citenamefont {Force}\ \emph {et~al.}(2006)\citenamefont {Force}, \citenamefont {Arnaud}, \citenamefont {Babak}, \citenamefont {Baker}, \citenamefont {Benacquista}, \citenamefont {Cornish}, \citenamefont {Cutler}, \citenamefont {Larson}, \citenamefont {Sathyaprakash}, \citenamefont {Vallisneri} \emph {et~al.}}]{mock2006overview}%
  \BibitemOpen
  \bibfield  {author} {\bibinfo {author} {\bibfnamefont {M.~L. D. C.~T.}\ \bibnamefont {Force}}, \bibinfo {author} {\bibfnamefont {K.~A.}\ \bibnamefont {Arnaud}}, \bibinfo {author} {\bibfnamefont {S.}~\bibnamefont {Babak}}, \bibinfo {author} {\bibfnamefont {J.~G.}\ \bibnamefont {Baker}}, \bibinfo {author} {\bibfnamefont {M.~J.}\ \bibnamefont {Benacquista}}, \bibinfo {author} {\bibfnamefont {N.~J.}\ \bibnamefont {Cornish}}, \bibinfo {author} {\bibfnamefont {C.}~\bibnamefont {Cutler}}, \bibinfo {author} {\bibfnamefont {S.~L.}\ \bibnamefont {Larson}}, \bibinfo {author} {\bibfnamefont {B.}~\bibnamefont {Sathyaprakash}}, \bibinfo {author} {\bibfnamefont {M.}~\bibnamefont {Vallisneri}}, \emph {et~al.},\ }\bibfield  {title} {\bibinfo {title} {An overview of the mock lisa data challenges},\ }in\ \href@noop {} {\emph {\bibinfo {booktitle} {AIP Conference Proceedings}}},\ Vol.\ \bibinfo {volume} {873}\ (\bibinfo {organization} {American Institute of Physics},\ \bibinfo {year} {2006})\ pp.\ \bibinfo {pages}
  {619--624}\BibitemShut {NoStop}%
\bibitem [{\citenamefont {Katz}(2022)}]{katz2022fully}%
  \BibitemOpen
  \bibfield  {author} {\bibinfo {author} {\bibfnamefont {M.~L.}\ \bibnamefont {Katz}},\ }\bibfield  {title} {\bibinfo {title} {Fully automated end-to-end pipeline for massive black hole binary signal extraction from lisa data},\ }\href@noop {} {\bibfield  {journal} {\bibinfo  {journal} {Physical Review D}\ }\textbf {\bibinfo {volume} {105}},\ \bibinfo {pages} {044055} (\bibinfo {year} {2022})}\BibitemShut {NoStop}%
\bibitem [{\citenamefont {Vallisneri}(2005)}]{vallisneri2005synthetic}%
  \BibitemOpen
  \bibfield  {author} {\bibinfo {author} {\bibfnamefont {M.}~\bibnamefont {Vallisneri}},\ }\bibfield  {title} {\bibinfo {title} {Synthetic lisa: Simulating time delay interferometry in a model lisa},\ }\href@noop {} {\bibfield  {journal} {\bibinfo  {journal} {Physical Review D}\ }\textbf {\bibinfo {volume} {71}},\ \bibinfo {pages} {022001} (\bibinfo {year} {2005})}\BibitemShut {NoStop}%
\bibitem [{\citenamefont {Finn}(1992)}]{finn1992detection}%
  \BibitemOpen
  \bibfield  {author} {\bibinfo {author} {\bibfnamefont {L.~S.}\ \bibnamefont {Finn}},\ }\bibfield  {title} {\bibinfo {title} {Detection, measurement, and gravitational radiation},\ }\href@noop {} {\bibfield  {journal} {\bibinfo  {journal} {Physical Review D}\ }\textbf {\bibinfo {volume} {46}},\ \bibinfo {pages} {5236} (\bibinfo {year} {1992})}\BibitemShut {NoStop}%
\bibitem [{\citenamefont {Usman}\ \emph {et~al.}(2016)\citenamefont {Usman}, \citenamefont {Nitz}, \citenamefont {Harry}, \citenamefont {Biwer}, \citenamefont {Brown}, \citenamefont {Cabero}, \citenamefont {Capano}, \citenamefont {Dal~Canton}, \citenamefont {Dent}, \citenamefont {Fairhurst} \emph {et~al.}}]{usman2016pycbc}%
  \BibitemOpen
  \bibfield  {author} {\bibinfo {author} {\bibfnamefont {S.~A.}\ \bibnamefont {Usman}}, \bibinfo {author} {\bibfnamefont {A.~H.}\ \bibnamefont {Nitz}}, \bibinfo {author} {\bibfnamefont {I.~W.}\ \bibnamefont {Harry}}, \bibinfo {author} {\bibfnamefont {C.~M.}\ \bibnamefont {Biwer}}, \bibinfo {author} {\bibfnamefont {D.~A.}\ \bibnamefont {Brown}}, \bibinfo {author} {\bibfnamefont {M.}~\bibnamefont {Cabero}}, \bibinfo {author} {\bibfnamefont {C.~D.}\ \bibnamefont {Capano}}, \bibinfo {author} {\bibfnamefont {T.}~\bibnamefont {Dal~Canton}}, \bibinfo {author} {\bibfnamefont {T.}~\bibnamefont {Dent}}, \bibinfo {author} {\bibfnamefont {S.}~\bibnamefont {Fairhurst}}, \emph {et~al.},\ }\bibfield  {title} {\bibinfo {title} {The pycbc search for gravitational waves from compact binary coalescence},\ }\href@noop {} {\bibfield  {journal} {\bibinfo  {journal} {Classical and Quantum Gravity}\ }\textbf {\bibinfo {volume} {33}},\ \bibinfo {pages} {215004} (\bibinfo {year} {2016})}\BibitemShut {NoStop}%
\bibitem [{\citenamefont {Zhao}\ \emph {et~al.}(2023)\citenamefont {Zhao}, \citenamefont {Lyu}, \citenamefont {Wang}, \citenamefont {Cao},\ and\ \citenamefont {Ren}}]{zhao2023space}%
  \BibitemOpen
  \bibfield  {author} {\bibinfo {author} {\bibfnamefont {T.}~\bibnamefont {Zhao}}, \bibinfo {author} {\bibfnamefont {R.}~\bibnamefont {Lyu}}, \bibinfo {author} {\bibfnamefont {H.}~\bibnamefont {Wang}}, \bibinfo {author} {\bibfnamefont {Z.}~\bibnamefont {Cao}},\ and\ \bibinfo {author} {\bibfnamefont {Z.}~\bibnamefont {Ren}},\ }\bibfield  {title} {\bibinfo {title} {Space-based gravitational wave signal detection and extraction with deep neural network},\ }\href@noop {} {\bibfield  {journal} {\bibinfo  {journal} {Communications Physics}\ }\textbf {\bibinfo {volume} {6}},\ \bibinfo {pages} {212} (\bibinfo {year} {2023})}\BibitemShut {NoStop}%
\bibitem [{\citenamefont {Wang}\ \emph {et~al.}(2020{\natexlab{a}})\citenamefont {Wang}, \citenamefont {Wu}, \citenamefont {Cao}, \citenamefont {Liu},\ and\ \citenamefont {Zhu}}]{wang2020gravitational}%
  \BibitemOpen
  \bibfield  {author} {\bibinfo {author} {\bibfnamefont {H.}~\bibnamefont {Wang}}, \bibinfo {author} {\bibfnamefont {S.}~\bibnamefont {Wu}}, \bibinfo {author} {\bibfnamefont {Z.}~\bibnamefont {Cao}}, \bibinfo {author} {\bibfnamefont {X.}~\bibnamefont {Liu}},\ and\ \bibinfo {author} {\bibfnamefont {J.-Y.}\ \bibnamefont {Zhu}},\ }\bibfield  {title} {\bibinfo {title} {Gravitational-wave signal recognition of ligo data by deep learning},\ }\href@noop {} {\bibfield  {journal} {\bibinfo  {journal} {Physical Review D}\ }\textbf {\bibinfo {volume} {101}},\ \bibinfo {pages} {104003} (\bibinfo {year} {2020}{\natexlab{a}})}\BibitemShut {NoStop}%
\bibitem [{\citenamefont {De~Luca}\ \emph {et~al.}(2020)\citenamefont {De~Luca}, \citenamefont {Franciolini}, \citenamefont {Pani},\ and\ \citenamefont {Riotto}}]{de2020primordial}%
  \BibitemOpen
  \bibfield  {author} {\bibinfo {author} {\bibfnamefont {V.}~\bibnamefont {De~Luca}}, \bibinfo {author} {\bibfnamefont {G.}~\bibnamefont {Franciolini}}, \bibinfo {author} {\bibfnamefont {P.}~\bibnamefont {Pani}},\ and\ \bibinfo {author} {\bibfnamefont {A.}~\bibnamefont {Riotto}},\ }\bibfield  {title} {\bibinfo {title} {Primordial black holes confront ligo/virgo data: current situation},\ }\href@noop {} {\bibfield  {journal} {\bibinfo  {journal} {Journal of Cosmology and Astroparticle Physics}\ }\textbf {\bibinfo {volume} {2020}}\bibinfo  {number} { (06)},\ \bibinfo {pages} {044}}\BibitemShut {NoStop}%
\bibitem [{\citenamefont {Chen}\ \emph {et~al.}(2022)\citenamefont {Chen}, \citenamefont {Cazenave}, \citenamefont {Dahle}, \citenamefont {Llovel}, \citenamefont {Panet}, \citenamefont {Pfeffer},\ and\ \citenamefont {Moreira}}]{chen2022applications}%
  \BibitemOpen
\bibfield  {number} {  }\bibfield  {author} {\bibinfo {author} {\bibfnamefont {J.}~\bibnamefont {Chen}}, \bibinfo {author} {\bibfnamefont {A.}~\bibnamefont {Cazenave}}, \bibinfo {author} {\bibfnamefont {C.}~\bibnamefont {Dahle}}, \bibinfo {author} {\bibfnamefont {W.}~\bibnamefont {Llovel}}, \bibinfo {author} {\bibfnamefont {I.}~\bibnamefont {Panet}}, \bibinfo {author} {\bibfnamefont {J.}~\bibnamefont {Pfeffer}},\ and\ \bibinfo {author} {\bibfnamefont {L.}~\bibnamefont {Moreira}},\ }\bibfield  {title} {\bibinfo {title} {Applications and challenges of grace and grace follow-on satellite gravimetry},\ }\href@noop {} {\bibfield  {journal} {\bibinfo  {journal} {Surveys in Geophysics}\ ,\ \bibinfo {pages} {1}} (\bibinfo {year} {2022})}\BibitemShut {NoStop}%
\bibitem [{\citenamefont {Armano}\ \emph {et~al.}(2009)\citenamefont {Armano}, \citenamefont {Benedetti}, \citenamefont {Bogenstahl}, \citenamefont {Bortoluzzi}, \citenamefont {Bosetti}, \citenamefont {Brandt}, \citenamefont {Cavalleri}, \citenamefont {Ciani}, \citenamefont {Cristofolini}, \citenamefont {Cruise} \emph {et~al.}}]{armano2009lisa}%
  \BibitemOpen
  \bibfield  {author} {\bibinfo {author} {\bibfnamefont {M.}~\bibnamefont {Armano}}, \bibinfo {author} {\bibfnamefont {M.}~\bibnamefont {Benedetti}}, \bibinfo {author} {\bibfnamefont {J.}~\bibnamefont {Bogenstahl}}, \bibinfo {author} {\bibfnamefont {D.}~\bibnamefont {Bortoluzzi}}, \bibinfo {author} {\bibfnamefont {P.}~\bibnamefont {Bosetti}}, \bibinfo {author} {\bibfnamefont {N.}~\bibnamefont {Brandt}}, \bibinfo {author} {\bibfnamefont {A.}~\bibnamefont {Cavalleri}}, \bibinfo {author} {\bibfnamefont {G.}~\bibnamefont {Ciani}}, \bibinfo {author} {\bibfnamefont {I.}~\bibnamefont {Cristofolini}}, \bibinfo {author} {\bibfnamefont {A.}~\bibnamefont {Cruise}}, \emph {et~al.},\ }\bibfield  {title} {\bibinfo {title} {Lisa pathfinder: the experiment and the route to lisa},\ }\href@noop {} {\bibfield  {journal} {\bibinfo  {journal} {Classical and Quantum Gravity}\ }\textbf {\bibinfo {volume} {26}},\ \bibinfo {pages} {094001} (\bibinfo {year} {2009})}\BibitemShut {NoStop}%
\bibitem [{\citenamefont {Collaboration}\ \emph {et~al.}(2021)\citenamefont {Collaboration}, \citenamefont {Wu}, \citenamefont {Luo}, \citenamefont {Wang}, \citenamefont {Bai}, \citenamefont {Bian}, \citenamefont {Cai}, \citenamefont {Cai}, \citenamefont {Cai}, \citenamefont {Cao} \emph {et~al.}}]{taiji2021taiji}%
  \BibitemOpen
  \bibfield  {author} {\bibinfo {author} {\bibfnamefont {T.~S.}\ \bibnamefont {Collaboration}}, \bibinfo {author} {\bibfnamefont {Y.-L.}\ \bibnamefont {Wu}}, \bibinfo {author} {\bibfnamefont {Z.-R.}\ \bibnamefont {Luo}}, \bibinfo {author} {\bibfnamefont {J.-Y.}\ \bibnamefont {Wang}}, \bibinfo {author} {\bibfnamefont {M.}~\bibnamefont {Bai}}, \bibinfo {author} {\bibfnamefont {W.}~\bibnamefont {Bian}}, \bibinfo {author} {\bibfnamefont {H.-W.}\ \bibnamefont {Cai}}, \bibinfo {author} {\bibfnamefont {R.-G.}\ \bibnamefont {Cai}}, \bibinfo {author} {\bibfnamefont {Z.-M.}\ \bibnamefont {Cai}}, \bibinfo {author} {\bibfnamefont {J.}~\bibnamefont {Cao}}, \emph {et~al.},\ }\href@noop {} {\bibinfo {title} {Taiji program in space for gravitational universe with the first run key technologies test in taiji-1}} (\bibinfo {year} {2021})\BibitemShut {NoStop}%
\bibitem [{\citenamefont {Baghi}\ \emph {et~al.}(2019)\citenamefont {Baghi}, \citenamefont {Thorpe}, \citenamefont {Slutsky}, \citenamefont {Baker}, \citenamefont {Dal~Canton}, \citenamefont {Korsakova},\ and\ \citenamefont {Karnesis}}]{baghi2019gravitational}%
  \BibitemOpen
  \bibfield  {author} {\bibinfo {author} {\bibfnamefont {Q.}~\bibnamefont {Baghi}}, \bibinfo {author} {\bibfnamefont {J.~I.}\ \bibnamefont {Thorpe}}, \bibinfo {author} {\bibfnamefont {J.}~\bibnamefont {Slutsky}}, \bibinfo {author} {\bibfnamefont {J.}~\bibnamefont {Baker}}, \bibinfo {author} {\bibfnamefont {T.}~\bibnamefont {Dal~Canton}}, \bibinfo {author} {\bibfnamefont {N.}~\bibnamefont {Korsakova}},\ and\ \bibinfo {author} {\bibfnamefont {N.}~\bibnamefont {Karnesis}},\ }\bibfield  {title} {\bibinfo {title} {Gravitational-wave parameter estimation with gaps in lisa: a bayesian data augmentation method},\ }\href@noop {} {\bibfield  {journal} {\bibinfo  {journal} {Physical Review D}\ }\textbf {\bibinfo {volume} {100}},\ \bibinfo {pages} {022003} (\bibinfo {year} {2019})}\BibitemShut {NoStop}%
\bibitem [{\citenamefont {Armano}\ \emph {et~al.}(2018)\citenamefont {Armano}, \citenamefont {Audley}, \citenamefont {Baird}, \citenamefont {Binetruy}, \citenamefont {Born}, \citenamefont {Bortoluzzi}, \citenamefont {Castelli}, \citenamefont {Cavalleri}, \citenamefont {Cesarini}, \citenamefont {Cruise} \emph {et~al.}}]{armano2018beyond}%
  \BibitemOpen
  \bibfield  {author} {\bibinfo {author} {\bibfnamefont {M.}~\bibnamefont {Armano}}, \bibinfo {author} {\bibfnamefont {H.}~\bibnamefont {Audley}}, \bibinfo {author} {\bibfnamefont {J.}~\bibnamefont {Baird}}, \bibinfo {author} {\bibfnamefont {P.}~\bibnamefont {Binetruy}}, \bibinfo {author} {\bibfnamefont {M.}~\bibnamefont {Born}}, \bibinfo {author} {\bibfnamefont {D.}~\bibnamefont {Bortoluzzi}}, \bibinfo {author} {\bibfnamefont {E.}~\bibnamefont {Castelli}}, \bibinfo {author} {\bibfnamefont {A.}~\bibnamefont {Cavalleri}}, \bibinfo {author} {\bibfnamefont {A.}~\bibnamefont {Cesarini}}, \bibinfo {author} {\bibfnamefont {A.}~\bibnamefont {Cruise}}, \emph {et~al.},\ }\bibfield  {title} {\bibinfo {title} {Beyond the required lisa free-fall performance: new lisa pathfinder results down to 20 $\mu$ hz},\ }\href@noop {} {\bibfield  {journal} {\bibinfo  {journal} {Physical review letters}\ }\textbf {\bibinfo {volume} {120}},\ \bibinfo {pages} {061101} (\bibinfo {year} {2018})}\BibitemShut {NoStop}%
\bibitem [{\citenamefont {Robson}\ and\ \citenamefont {Cornish}(2019)}]{robson2019detecting}%
  \BibitemOpen
  \bibfield  {author} {\bibinfo {author} {\bibfnamefont {T.}~\bibnamefont {Robson}}\ and\ \bibinfo {author} {\bibfnamefont {N.~J.}\ \bibnamefont {Cornish}},\ }\bibfield  {title} {\bibinfo {title} {Detecting gravitational wave bursts with lisa in the presence of instrumental glitches},\ }\href@noop {} {\bibfield  {journal} {\bibinfo  {journal} {Physical Review D}\ }\textbf {\bibinfo {volume} {99}},\ \bibinfo {pages} {024019} (\bibinfo {year} {2019})}\BibitemShut {NoStop}%
\bibitem [{\citenamefont {Armano}\ \emph {et~al.}(2022)\citenamefont {Armano}, \citenamefont {Audley}, \citenamefont {Baird}, \citenamefont {Binetruy}, \citenamefont {Born}, \citenamefont {Bortoluzzi}, \citenamefont {Castelli}, \citenamefont {Cavalleri}, \citenamefont {Cesarini}, \citenamefont {Chiavegato} \emph {et~al.}}]{armano2022transient}%
  \BibitemOpen
  \bibfield  {author} {\bibinfo {author} {\bibfnamefont {M.}~\bibnamefont {Armano}}, \bibinfo {author} {\bibfnamefont {H.}~\bibnamefont {Audley}}, \bibinfo {author} {\bibfnamefont {J.}~\bibnamefont {Baird}}, \bibinfo {author} {\bibfnamefont {P.}~\bibnamefont {Binetruy}}, \bibinfo {author} {\bibfnamefont {M.}~\bibnamefont {Born}}, \bibinfo {author} {\bibfnamefont {D.}~\bibnamefont {Bortoluzzi}}, \bibinfo {author} {\bibfnamefont {E.}~\bibnamefont {Castelli}}, \bibinfo {author} {\bibfnamefont {A.}~\bibnamefont {Cavalleri}}, \bibinfo {author} {\bibfnamefont {A.}~\bibnamefont {Cesarini}}, \bibinfo {author} {\bibfnamefont {V.}~\bibnamefont {Chiavegato}}, \emph {et~al.},\ }\bibfield  {title} {\bibinfo {title} {Transient acceleration events in lisa pathfinder data: Properties and possible physical origin},\ }\href@noop {} {\bibfield  {journal} {\bibinfo  {journal} {Physical Review D}\ }\textbf {\bibinfo {volume} {106}},\ \bibinfo {pages} {062001} (\bibinfo {year} {2022})}\BibitemShut {NoStop}%
\bibitem [{\citenamefont {Armano}\ \emph {et~al.}(2016)\citenamefont {Armano}, \citenamefont {Audley}, \citenamefont {Auger}, \citenamefont {Baird}, \citenamefont {Bassan}, \citenamefont {Binetruy}, \citenamefont {Born}, \citenamefont {Bortoluzzi}, \citenamefont {Brandt}, \citenamefont {Caleno} \emph {et~al.}}]{armano2016sub}%
  \BibitemOpen
  \bibfield  {author} {\bibinfo {author} {\bibfnamefont {M.}~\bibnamefont {Armano}}, \bibinfo {author} {\bibfnamefont {H.}~\bibnamefont {Audley}}, \bibinfo {author} {\bibfnamefont {G.}~\bibnamefont {Auger}}, \bibinfo {author} {\bibfnamefont {J.~T.}\ \bibnamefont {Baird}}, \bibinfo {author} {\bibfnamefont {M.}~\bibnamefont {Bassan}}, \bibinfo {author} {\bibfnamefont {P.}~\bibnamefont {Binetruy}}, \bibinfo {author} {\bibfnamefont {M.}~\bibnamefont {Born}}, \bibinfo {author} {\bibfnamefont {D.}~\bibnamefont {Bortoluzzi}}, \bibinfo {author} {\bibfnamefont {N.}~\bibnamefont {Brandt}}, \bibinfo {author} {\bibfnamefont {M.}~\bibnamefont {Caleno}}, \emph {et~al.},\ }\bibfield  {title} {\bibinfo {title} {Sub-femto-g free fall for space-based gravitational wave observatories: Lisa pathfinder results},\ }\href@noop {} {\bibfield  {journal} {\bibinfo  {journal} {Physical review letters}\ }\textbf {\bibinfo {volume} {116}},\ \bibinfo {pages} {231101} (\bibinfo {year} {2016})}\BibitemShut {NoStop}%
\bibitem [{\citenamefont {Frommknecht}(2007)}]{frommknecht2007integrated}%
  \BibitemOpen
  \bibfield  {author} {\bibinfo {author} {\bibfnamefont {B.}~\bibnamefont {Frommknecht}},\ }\emph {\bibinfo {title} {Integrated sensor analysis of the GRACE mission}},\ \href@noop {} {Ph.D. thesis},\ \bibinfo  {school} {Technische Universit{\"a}t M{\"u}nchen} (\bibinfo {year} {2007})\BibitemShut {NoStop}%
\bibitem [{\citenamefont {Sheard}\ \emph {et~al.}(2012)\citenamefont {Sheard}, \citenamefont {Heinzel}, \citenamefont {Danzmann}, \citenamefont {Shaddock}, \citenamefont {Klipstein},\ and\ \citenamefont {Folkner}}]{sheard2012intersatellite}%
  \BibitemOpen
  \bibfield  {author} {\bibinfo {author} {\bibfnamefont {B.}~\bibnamefont {Sheard}}, \bibinfo {author} {\bibfnamefont {G.}~\bibnamefont {Heinzel}}, \bibinfo {author} {\bibfnamefont {K.}~\bibnamefont {Danzmann}}, \bibinfo {author} {\bibfnamefont {D.}~\bibnamefont {Shaddock}}, \bibinfo {author} {\bibfnamefont {W.}~\bibnamefont {Klipstein}},\ and\ \bibinfo {author} {\bibfnamefont {W.}~\bibnamefont {Folkner}},\ }\bibfield  {title} {\bibinfo {title} {Intersatellite laser ranging instrument for the grace follow-on mission},\ }\href@noop {} {\bibfield  {journal} {\bibinfo  {journal} {Journal of Geodesy}\ }\textbf {\bibinfo {volume} {86}},\ \bibinfo {pages} {1083} (\bibinfo {year} {2012})}\BibitemShut {NoStop}%
\bibitem [{tai(2021)}]{taiji2021china}%
  \BibitemOpen
  \bibfield  {title} {\bibinfo {title} {China’s first step towards probing the expanding universe and the nature of gravity using a space borne gravitational wave antenna},\ }\href@noop {} {\bibfield  {journal} {\bibinfo  {journal} {Communications Physics}\ }\textbf {\bibinfo {volume} {4}},\ \bibinfo {pages} {34} (\bibinfo {year} {2021})}\BibitemShut {NoStop}%
\bibitem [{\citenamefont {Spadaro}\ \emph {et~al.}(2023)\citenamefont {Spadaro}, \citenamefont {Buscicchio}, \citenamefont {Vetrugno}, \citenamefont {Klein}, \citenamefont {Gerosa}, \citenamefont {Vitale}, \citenamefont {Dolesi}, \citenamefont {Weber},\ and\ \citenamefont {Colpi}}]{spadaro2023glitch}%
  \BibitemOpen
  \bibfield  {author} {\bibinfo {author} {\bibfnamefont {A.}~\bibnamefont {Spadaro}}, \bibinfo {author} {\bibfnamefont {R.}~\bibnamefont {Buscicchio}}, \bibinfo {author} {\bibfnamefont {D.}~\bibnamefont {Vetrugno}}, \bibinfo {author} {\bibfnamefont {A.}~\bibnamefont {Klein}}, \bibinfo {author} {\bibfnamefont {D.}~\bibnamefont {Gerosa}}, \bibinfo {author} {\bibfnamefont {S.}~\bibnamefont {Vitale}}, \bibinfo {author} {\bibfnamefont {R.}~\bibnamefont {Dolesi}}, \bibinfo {author} {\bibfnamefont {W.~J.}\ \bibnamefont {Weber}},\ and\ \bibinfo {author} {\bibfnamefont {M.}~\bibnamefont {Colpi}},\ }\bibfield  {title} {\bibinfo {title} {Glitch systematics on the observation of massive black-hole binaries with lisa},\ }\href@noop {} {\bibfield  {journal} {\bibinfo  {journal} {arXiv preprint arXiv:2306.03923}\ } (\bibinfo {year} {2023})}\BibitemShut {NoStop}%
\bibitem [{\citenamefont {George}\ and\ \citenamefont {Huerta}(2018)}]{george2018deep}%
  \BibitemOpen
  \bibfield  {author} {\bibinfo {author} {\bibfnamefont {D.}~\bibnamefont {George}}\ and\ \bibinfo {author} {\bibfnamefont {E.}~\bibnamefont {Huerta}},\ }\bibfield  {title} {\bibinfo {title} {Deep neural networks to enable real-time multimessenger astrophysics},\ }\href@noop {} {\bibfield  {journal} {\bibinfo  {journal} {Physical Review D}\ }\textbf {\bibinfo {volume} {97}},\ \bibinfo {pages} {044039} (\bibinfo {year} {2018})}\BibitemShut {NoStop}%
\bibitem [{\citenamefont {Gabbard}\ \emph {et~al.}(2018)\citenamefont {Gabbard}, \citenamefont {Williams}, \citenamefont {Hayes},\ and\ \citenamefont {Messenger}}]{gabbard2018matching}%
  \BibitemOpen
  \bibfield  {author} {\bibinfo {author} {\bibfnamefont {H.}~\bibnamefont {Gabbard}}, \bibinfo {author} {\bibfnamefont {M.}~\bibnamefont {Williams}}, \bibinfo {author} {\bibfnamefont {F.}~\bibnamefont {Hayes}},\ and\ \bibinfo {author} {\bibfnamefont {C.}~\bibnamefont {Messenger}},\ }\bibfield  {title} {\bibinfo {title} {Matching matched filtering with deep networks for gravitational-wave astronomy},\ }\href@noop {} {\bibfield  {journal} {\bibinfo  {journal} {Physical review letters}\ }\textbf {\bibinfo {volume} {120}},\ \bibinfo {pages} {141103} (\bibinfo {year} {2018})}\BibitemShut {NoStop}%
\bibitem [{\citenamefont {Gabbard}\ \emph {et~al.}(2022)\citenamefont {Gabbard}, \citenamefont {Messenger}, \citenamefont {Heng}, \citenamefont {Tonolini},\ and\ \citenamefont {Murray-Smith}}]{gabbard2022bayesian}%
  \BibitemOpen
  \bibfield  {author} {\bibinfo {author} {\bibfnamefont {H.}~\bibnamefont {Gabbard}}, \bibinfo {author} {\bibfnamefont {C.}~\bibnamefont {Messenger}}, \bibinfo {author} {\bibfnamefont {I.~S.}\ \bibnamefont {Heng}}, \bibinfo {author} {\bibfnamefont {F.}~\bibnamefont {Tonolini}},\ and\ \bibinfo {author} {\bibfnamefont {R.}~\bibnamefont {Murray-Smith}},\ }\bibfield  {title} {\bibinfo {title} {Bayesian parameter estimation using conditional variational autoencoders for gravitational-wave astronomy},\ }\href@noop {} {\bibfield  {journal} {\bibinfo  {journal} {Nature Physics}\ }\textbf {\bibinfo {volume} {18}},\ \bibinfo {pages} {112} (\bibinfo {year} {2022})}\BibitemShut {NoStop}%
\bibitem [{\citenamefont {Dax}\ \emph {et~al.}(2021)\citenamefont {Dax}, \citenamefont {Green}, \citenamefont {Gair}, \citenamefont {Macke}, \citenamefont {Buonanno},\ and\ \citenamefont {Sch{\"o}lkopf}}]{dax2021real}%
  \BibitemOpen
  \bibfield  {author} {\bibinfo {author} {\bibfnamefont {M.}~\bibnamefont {Dax}}, \bibinfo {author} {\bibfnamefont {S.~R.}\ \bibnamefont {Green}}, \bibinfo {author} {\bibfnamefont {J.}~\bibnamefont {Gair}}, \bibinfo {author} {\bibfnamefont {J.~H.}\ \bibnamefont {Macke}}, \bibinfo {author} {\bibfnamefont {A.}~\bibnamefont {Buonanno}},\ and\ \bibinfo {author} {\bibfnamefont {B.}~\bibnamefont {Sch{\"o}lkopf}},\ }\bibfield  {title} {\bibinfo {title} {Real-time gravitational wave science with neural posterior estimation},\ }\href@noop {} {\bibfield  {journal} {\bibinfo  {journal} {Physical review letters}\ }\textbf {\bibinfo {volume} {127}},\ \bibinfo {pages} {241103} (\bibinfo {year} {2021})}\BibitemShut {NoStop}%
\bibitem [{\citenamefont {Colgan}\ \emph {et~al.}(2020)\citenamefont {Colgan}, \citenamefont {Corley}, \citenamefont {Lau}, \citenamefont {Bartos}, \citenamefont {Wright}, \citenamefont {M{\'a}rka},\ and\ \citenamefont {M{\'a}rka}}]{colgan2020efficient}%
  \BibitemOpen
  \bibfield  {author} {\bibinfo {author} {\bibfnamefont {R.~E.}\ \bibnamefont {Colgan}}, \bibinfo {author} {\bibfnamefont {K.~R.}\ \bibnamefont {Corley}}, \bibinfo {author} {\bibfnamefont {Y.}~\bibnamefont {Lau}}, \bibinfo {author} {\bibfnamefont {I.}~\bibnamefont {Bartos}}, \bibinfo {author} {\bibfnamefont {J.~N.}\ \bibnamefont {Wright}}, \bibinfo {author} {\bibfnamefont {Z.}~\bibnamefont {M{\'a}rka}},\ and\ \bibinfo {author} {\bibfnamefont {S.}~\bibnamefont {M{\'a}rka}},\ }\bibfield  {title} {\bibinfo {title} {Efficient gravitational-wave glitch identification from environmental data through machine learning},\ }\href@noop {} {\bibfield  {journal} {\bibinfo  {journal} {Physical Review D}\ }\textbf {\bibinfo {volume} {101}},\ \bibinfo {pages} {102003} (\bibinfo {year} {2020})}\BibitemShut {NoStop}%
\bibitem [{\citenamefont {Cavaglia}\ \emph {et~al.}(2019)\citenamefont {Cavaglia}, \citenamefont {Staats},\ and\ \citenamefont {Gill}}]{cavaglia2019finding}%
  \BibitemOpen
  \bibfield  {author} {\bibinfo {author} {\bibfnamefont {M.}~\bibnamefont {Cavaglia}}, \bibinfo {author} {\bibfnamefont {K.}~\bibnamefont {Staats}},\ and\ \bibinfo {author} {\bibfnamefont {T.}~\bibnamefont {Gill}},\ }\bibfield  {title} {\bibinfo {title} {Finding the origin of noise transients in ligo data with machine learning},\ }\href@noop {} {\bibfield  {journal} {\bibinfo  {journal} {Communications in Computational Physics}\ }\textbf {\bibinfo {volume} {25}} (\bibinfo {year} {2019})}\BibitemShut {NoStop}%
\bibitem [{\citenamefont {Razzano}\ and\ \citenamefont {Cuoco}(2018)}]{razzano2018image}%
  \BibitemOpen
  \bibfield  {author} {\bibinfo {author} {\bibfnamefont {M.}~\bibnamefont {Razzano}}\ and\ \bibinfo {author} {\bibfnamefont {E.}~\bibnamefont {Cuoco}},\ }\bibfield  {title} {\bibinfo {title} {Image-based deep learning for classification of noise transients in gravitational wave detectors},\ }\href@noop {} {\bibfield  {journal} {\bibinfo  {journal} {Classical and Quantum Gravity}\ }\textbf {\bibinfo {volume} {35}},\ \bibinfo {pages} {095016} (\bibinfo {year} {2018})}\BibitemShut {NoStop}%
\bibitem [{\citenamefont {Ormiston}\ \emph {et~al.}(2020)\citenamefont {Ormiston}, \citenamefont {Nguyen}, \citenamefont {Coughlin}, \citenamefont {Adhikari},\ and\ \citenamefont {Katsavounidis}}]{ormiston2020noise}%
  \BibitemOpen
  \bibfield  {author} {\bibinfo {author} {\bibfnamefont {R.}~\bibnamefont {Ormiston}}, \bibinfo {author} {\bibfnamefont {T.}~\bibnamefont {Nguyen}}, \bibinfo {author} {\bibfnamefont {M.}~\bibnamefont {Coughlin}}, \bibinfo {author} {\bibfnamefont {R.~X.}\ \bibnamefont {Adhikari}},\ and\ \bibinfo {author} {\bibfnamefont {E.}~\bibnamefont {Katsavounidis}},\ }\bibfield  {title} {\bibinfo {title} {Noise reduction in gravitational-wave data via deep learning},\ }\href@noop {} {\bibfield  {journal} {\bibinfo  {journal} {Physical Review Research}\ }\textbf {\bibinfo {volume} {2}},\ \bibinfo {pages} {033066} (\bibinfo {year} {2020})}\BibitemShut {NoStop}%
\bibitem [{\citenamefont {Torres-Forne}\ \emph {et~al.}(2016)\citenamefont {Torres-Forne}, \citenamefont {Marquina}, \citenamefont {Font},\ and\ \citenamefont {Ibanez}}]{torres2016denoising}%
  \BibitemOpen
  \bibfield  {author} {\bibinfo {author} {\bibfnamefont {A.}~\bibnamefont {Torres-Forne}}, \bibinfo {author} {\bibfnamefont {A.}~\bibnamefont {Marquina}}, \bibinfo {author} {\bibfnamefont {J.~A.}\ \bibnamefont {Font}},\ and\ \bibinfo {author} {\bibfnamefont {J.~M.}\ \bibnamefont {Ibanez}},\ }\bibfield  {title} {\bibinfo {title} {Denoising of gravitational wave signals via dictionary learning algorithms},\ }\href@noop {} {\bibfield  {journal} {\bibinfo  {journal} {Physical Review D}\ }\textbf {\bibinfo {volume} {94}},\ \bibinfo {pages} {124040} (\bibinfo {year} {2016})}\BibitemShut {NoStop}%
\bibitem [{\citenamefont {Wei}\ and\ \citenamefont {Huerta}(2020)}]{wei2020gravitational}%
  \BibitemOpen
  \bibfield  {author} {\bibinfo {author} {\bibfnamefont {W.}~\bibnamefont {Wei}}\ and\ \bibinfo {author} {\bibfnamefont {E.}~\bibnamefont {Huerta}},\ }\bibfield  {title} {\bibinfo {title} {Gravitational wave denoising of binary black hole mergers with deep learning},\ }\href@noop {} {\bibfield  {journal} {\bibinfo  {journal} {Physics Letters B}\ }\textbf {\bibinfo {volume} {800}},\ \bibinfo {pages} {135081} (\bibinfo {year} {2020})}\BibitemShut {NoStop}%
\bibitem [{\citenamefont {Cannon}\ \emph {et~al.}(2021)\citenamefont {Cannon}, \citenamefont {Caudill}, \citenamefont {Chan}, \citenamefont {Cousins}, \citenamefont {Creighton}, \citenamefont {Ewing}, \citenamefont {Fong}, \citenamefont {Godwin}, \citenamefont {Hanna}, \citenamefont {Hooper} \emph {et~al.}}]{cannon2021gstlal}%
  \BibitemOpen
  \bibfield  {author} {\bibinfo {author} {\bibfnamefont {K.}~\bibnamefont {Cannon}}, \bibinfo {author} {\bibfnamefont {S.}~\bibnamefont {Caudill}}, \bibinfo {author} {\bibfnamefont {C.}~\bibnamefont {Chan}}, \bibinfo {author} {\bibfnamefont {B.}~\bibnamefont {Cousins}}, \bibinfo {author} {\bibfnamefont {J.~D.}\ \bibnamefont {Creighton}}, \bibinfo {author} {\bibfnamefont {B.}~\bibnamefont {Ewing}}, \bibinfo {author} {\bibfnamefont {H.}~\bibnamefont {Fong}}, \bibinfo {author} {\bibfnamefont {P.}~\bibnamefont {Godwin}}, \bibinfo {author} {\bibfnamefont {C.}~\bibnamefont {Hanna}}, \bibinfo {author} {\bibfnamefont {S.}~\bibnamefont {Hooper}}, \emph {et~al.},\ }\bibfield  {title} {\bibinfo {title} {Gstlal: A software framework for gravitational wave discovery},\ }\href@noop {} {\bibfield  {journal} {\bibinfo  {journal} {SoftwareX}\ }\textbf {\bibinfo {volume} {14}},\ \bibinfo {pages} {100680} (\bibinfo {year} {2021})}\BibitemShut {NoStop}%
\bibitem [{\citenamefont {Gondara}(2016)}]{gondara2016medical}%
  \BibitemOpen
  \bibfield  {author} {\bibinfo {author} {\bibfnamefont {L.}~\bibnamefont {Gondara}},\ }\bibfield  {title} {\bibinfo {title} {Medical image denoising using convolutional denoising autoencoders},\ }in\ \href@noop {} {\emph {\bibinfo {booktitle} {2016 IEEE 16th international conference on data mining workshops (ICDMW)}}}\ (\bibinfo {organization} {IEEE},\ \bibinfo {year} {2016})\ pp.\ \bibinfo {pages} {241--246}\BibitemShut {NoStop}%
\bibitem [{\citenamefont {Xie}\ \emph {et~al.}(2012)\citenamefont {Xie}, \citenamefont {Xu},\ and\ \citenamefont {Chen}}]{xie2012image}%
  \BibitemOpen
  \bibfield  {author} {\bibinfo {author} {\bibfnamefont {J.}~\bibnamefont {Xie}}, \bibinfo {author} {\bibfnamefont {L.}~\bibnamefont {Xu}},\ and\ \bibinfo {author} {\bibfnamefont {E.}~\bibnamefont {Chen}},\ }\bibfield  {title} {\bibinfo {title} {Image denoising and inpainting with deep neural networks},\ }\href@noop {} {\bibfield  {journal} {\bibinfo  {journal} {Advances in neural information processing systems}\ }\textbf {\bibinfo {volume} {25}} (\bibinfo {year} {2012})}\BibitemShut {NoStop}%
\bibitem [{\citenamefont {Vincent}\ \emph {et~al.}(2010)\citenamefont {Vincent}, \citenamefont {Larochelle}, \citenamefont {Lajoie}, \citenamefont {Bengio}, \citenamefont {Manzagol},\ and\ \citenamefont {Bottou}}]{vincent2010stacked}%
  \BibitemOpen
  \bibfield  {author} {\bibinfo {author} {\bibfnamefont {P.}~\bibnamefont {Vincent}}, \bibinfo {author} {\bibfnamefont {H.}~\bibnamefont {Larochelle}}, \bibinfo {author} {\bibfnamefont {I.}~\bibnamefont {Lajoie}}, \bibinfo {author} {\bibfnamefont {Y.}~\bibnamefont {Bengio}}, \bibinfo {author} {\bibfnamefont {P.-A.}\ \bibnamefont {Manzagol}},\ and\ \bibinfo {author} {\bibfnamefont {L.}~\bibnamefont {Bottou}},\ }\bibfield  {title} {\bibinfo {title} {Stacked denoising autoencoders: Learning useful representations in a deep network with a local denoising criterion.},\ }\href@noop {} {\bibfield  {journal} {\bibinfo  {journal} {Journal of machine learning research}\ }\textbf {\bibinfo {volume} {11}} (\bibinfo {year} {2010})}\BibitemShut {NoStop}%
\bibitem [{\citenamefont {Creswell}\ and\ \citenamefont {Bharath}(2018)}]{creswell2018denoising}%
  \BibitemOpen
  \bibfield  {author} {\bibinfo {author} {\bibfnamefont {A.}~\bibnamefont {Creswell}}\ and\ \bibinfo {author} {\bibfnamefont {A.~A.}\ \bibnamefont {Bharath}},\ }\bibfield  {title} {\bibinfo {title} {Denoising adversarial autoencoders},\ }\href@noop {} {\bibfield  {journal} {\bibinfo  {journal} {IEEE transactions on neural networks and learning systems}\ }\textbf {\bibinfo {volume} {30}},\ \bibinfo {pages} {968} (\bibinfo {year} {2018})}\BibitemShut {NoStop}%
\bibitem [{\citenamefont {Chiang}\ \emph {et~al.}(2019)\citenamefont {Chiang}, \citenamefont {Hsieh}, \citenamefont {Fu}, \citenamefont {Hung}, \citenamefont {Tsao},\ and\ \citenamefont {Chien}}]{chiang2019noise}%
  \BibitemOpen
  \bibfield  {author} {\bibinfo {author} {\bibfnamefont {H.-T.}\ \bibnamefont {Chiang}}, \bibinfo {author} {\bibfnamefont {Y.-Y.}\ \bibnamefont {Hsieh}}, \bibinfo {author} {\bibfnamefont {S.-W.}\ \bibnamefont {Fu}}, \bibinfo {author} {\bibfnamefont {K.-H.}\ \bibnamefont {Hung}}, \bibinfo {author} {\bibfnamefont {Y.}~\bibnamefont {Tsao}},\ and\ \bibinfo {author} {\bibfnamefont {S.-Y.}\ \bibnamefont {Chien}},\ }\bibfield  {title} {\bibinfo {title} {Noise reduction in ecg signals using fully convolutional denoising autoencoders},\ }\href@noop {} {\bibfield  {journal} {\bibinfo  {journal} {Ieee Access}\ }\textbf {\bibinfo {volume} {7}},\ \bibinfo {pages} {60806} (\bibinfo {year} {2019})}\BibitemShut {NoStop}%
\bibitem [{\citenamefont {Saad}\ and\ \citenamefont {Chen}(2020)}]{saad2020deep}%
  \BibitemOpen
  \bibfield  {author} {\bibinfo {author} {\bibfnamefont {O.~M.}\ \bibnamefont {Saad}}\ and\ \bibinfo {author} {\bibfnamefont {Y.}~\bibnamefont {Chen}},\ }\bibfield  {title} {\bibinfo {title} {Deep denoising autoencoder for seismic random noise attenuation},\ }\href@noop {} {\bibfield  {journal} {\bibinfo  {journal} {Geophysics}\ }\textbf {\bibinfo {volume} {85}},\ \bibinfo {pages} {V367} (\bibinfo {year} {2020})}\BibitemShut {NoStop}%
\bibitem [{\citenamefont {Xia}\ \emph {et~al.}(2017)\citenamefont {Xia}, \citenamefont {Li}, \citenamefont {Liu}, \citenamefont {Xu},\ and\ \citenamefont {de~Silva}}]{xia2017intelligent}%
  \BibitemOpen
  \bibfield  {author} {\bibinfo {author} {\bibfnamefont {M.}~\bibnamefont {Xia}}, \bibinfo {author} {\bibfnamefont {T.}~\bibnamefont {Li}}, \bibinfo {author} {\bibfnamefont {L.}~\bibnamefont {Liu}}, \bibinfo {author} {\bibfnamefont {L.}~\bibnamefont {Xu}},\ and\ \bibinfo {author} {\bibfnamefont {C.~W.}\ \bibnamefont {de~Silva}},\ }\bibfield  {title} {\bibinfo {title} {Intelligent fault diagnosis approach with unsupervised feature learning by stacked denoising autoencoder},\ }\href@noop {} {\bibfield  {journal} {\bibinfo  {journal} {IET Science, Measurement \& Technology}\ }\textbf {\bibinfo {volume} {11}},\ \bibinfo {pages} {687} (\bibinfo {year} {2017})}\BibitemShut {NoStop}%
\bibitem [{\citenamefont {Dasan}\ and\ \citenamefont {Panneerselvam}(2021)}]{dasan2021novel}%
  \BibitemOpen
  \bibfield  {author} {\bibinfo {author} {\bibfnamefont {E.}~\bibnamefont {Dasan}}\ and\ \bibinfo {author} {\bibfnamefont {I.}~\bibnamefont {Panneerselvam}},\ }\bibfield  {title} {\bibinfo {title} {A novel dimensionality reduction approach for ecg signal via convolutional denoising autoencoder with lstm},\ }\href@noop {} {\bibfield  {journal} {\bibinfo  {journal} {Biomedical Signal Processing and Control}\ }\textbf {\bibinfo {volume} {63}},\ \bibinfo {pages} {102225} (\bibinfo {year} {2021})}\BibitemShut {NoStop}%
\bibitem [{\citenamefont {Araki}\ \emph {et~al.}(2015)\citenamefont {Araki}, \citenamefont {Hayashi}, \citenamefont {Delcroix}, \citenamefont {Fujimoto}, \citenamefont {Takeda},\ and\ \citenamefont {Nakatani}}]{araki2015exploring}%
  \BibitemOpen
  \bibfield  {author} {\bibinfo {author} {\bibfnamefont {S.}~\bibnamefont {Araki}}, \bibinfo {author} {\bibfnamefont {T.}~\bibnamefont {Hayashi}}, \bibinfo {author} {\bibfnamefont {M.}~\bibnamefont {Delcroix}}, \bibinfo {author} {\bibfnamefont {M.}~\bibnamefont {Fujimoto}}, \bibinfo {author} {\bibfnamefont {K.}~\bibnamefont {Takeda}},\ and\ \bibinfo {author} {\bibfnamefont {T.}~\bibnamefont {Nakatani}},\ }\bibfield  {title} {\bibinfo {title} {Exploring multi-channel features for denoising-autoencoder-based speech enhancement},\ }in\ \href@noop {} {\emph {\bibinfo {booktitle} {2015 IEEE International Conference on Acoustics, Speech and Signal Processing (ICASSP)}}}\ (\bibinfo {organization} {IEEE},\ \bibinfo {year} {2015})\ pp.\ \bibinfo {pages} {116--120}\BibitemShut {NoStop}%
\bibitem [{\citenamefont {Lai}\ \emph {et~al.}(2016)\citenamefont {Lai}, \citenamefont {Chen}, \citenamefont {Wang}, \citenamefont {Lu}, \citenamefont {Tsao},\ and\ \citenamefont {Lee}}]{lai2016deep}%
  \BibitemOpen
  \bibfield  {author} {\bibinfo {author} {\bibfnamefont {Y.-H.}\ \bibnamefont {Lai}}, \bibinfo {author} {\bibfnamefont {F.}~\bibnamefont {Chen}}, \bibinfo {author} {\bibfnamefont {S.-S.}\ \bibnamefont {Wang}}, \bibinfo {author} {\bibfnamefont {X.}~\bibnamefont {Lu}}, \bibinfo {author} {\bibfnamefont {Y.}~\bibnamefont {Tsao}},\ and\ \bibinfo {author} {\bibfnamefont {C.-H.}\ \bibnamefont {Lee}},\ }\bibfield  {title} {\bibinfo {title} {A deep denoising autoencoder approach to improving the intelligibility of vocoded speech in cochlear implant simulation},\ }\href@noop {} {\bibfield  {journal} {\bibinfo  {journal} {IEEE Transactions on Biomedical Engineering}\ }\textbf {\bibinfo {volume} {64}},\ \bibinfo {pages} {1568} (\bibinfo {year} {2016})}\BibitemShut {NoStop}%
\bibitem [{\citenamefont {Feng}\ \emph {et~al.}(2014)\citenamefont {Feng}, \citenamefont {Zhang},\ and\ \citenamefont {Glass}}]{feng2014speech}%
  \BibitemOpen
  \bibfield  {author} {\bibinfo {author} {\bibfnamefont {X.}~\bibnamefont {Feng}}, \bibinfo {author} {\bibfnamefont {Y.}~\bibnamefont {Zhang}},\ and\ \bibinfo {author} {\bibfnamefont {J.}~\bibnamefont {Glass}},\ }\bibfield  {title} {\bibinfo {title} {Speech feature denoising and dereverberation via deep autoencoders for noisy reverberant speech recognition},\ }in\ \href@noop {} {\emph {\bibinfo {booktitle} {2014 IEEE international conference on acoustics, speech and signal processing (ICASSP)}}}\ (\bibinfo {organization} {IEEE},\ \bibinfo {year} {2014})\ pp.\ \bibinfo {pages} {1759--1763}\BibitemShut {NoStop}%
\bibitem [{\citenamefont {Lu}\ \emph {et~al.}(2013)\citenamefont {Lu}, \citenamefont {Tsao}, \citenamefont {Matsuda},\ and\ \citenamefont {Hori}}]{lu2013speech}%
  \BibitemOpen
  \bibfield  {author} {\bibinfo {author} {\bibfnamefont {X.}~\bibnamefont {Lu}}, \bibinfo {author} {\bibfnamefont {Y.}~\bibnamefont {Tsao}}, \bibinfo {author} {\bibfnamefont {S.}~\bibnamefont {Matsuda}},\ and\ \bibinfo {author} {\bibfnamefont {C.}~\bibnamefont {Hori}},\ }\bibfield  {title} {\bibinfo {title} {Speech enhancement based on deep denoising autoencoder.},\ }in\ \href@noop {} {\emph {\bibinfo {booktitle} {Interspeech}}},\ Vol.\ \bibinfo {volume} {2013}\ (\bibinfo {year} {2013})\ pp.\ \bibinfo {pages} {436--440}\BibitemShut {NoStop}%
\bibitem [{\citenamefont {Chatterjee}\ \emph {et~al.}(2021)\citenamefont {Chatterjee}, \citenamefont {Wen}, \citenamefont {Diakogiannis},\ and\ \citenamefont {Vinsen}}]{chatterjee2021extraction}%
  \BibitemOpen
  \bibfield  {author} {\bibinfo {author} {\bibfnamefont {C.}~\bibnamefont {Chatterjee}}, \bibinfo {author} {\bibfnamefont {L.}~\bibnamefont {Wen}}, \bibinfo {author} {\bibfnamefont {F.}~\bibnamefont {Diakogiannis}},\ and\ \bibinfo {author} {\bibfnamefont {K.}~\bibnamefont {Vinsen}},\ }\bibfield  {title} {\bibinfo {title} {Extraction of binary black hole gravitational wave signals from detector data using deep learning},\ }\href@noop {} {\bibfield  {journal} {\bibinfo  {journal} {Physical Review D}\ }\textbf {\bibinfo {volume} {104}},\ \bibinfo {pages} {064046} (\bibinfo {year} {2021})}\BibitemShut {NoStop}%
\bibitem [{\citenamefont {Krizhevsky}\ \emph {et~al.}(2012)\citenamefont {Krizhevsky}, \citenamefont {Sutskever},\ and\ \citenamefont {Hinton}}]{krizhevsky2012imagenet}%
  \BibitemOpen
  \bibfield  {author} {\bibinfo {author} {\bibfnamefont {A.}~\bibnamefont {Krizhevsky}}, \bibinfo {author} {\bibfnamefont {I.}~\bibnamefont {Sutskever}},\ and\ \bibinfo {author} {\bibfnamefont {G.~E.}\ \bibnamefont {Hinton}},\ }\bibfield  {title} {\bibinfo {title} {Imagenet classification with deep convolutional neural networks},\ }\href@noop {} {\bibfield  {journal} {\bibinfo  {journal} {Advances in neural information processing systems}\ }\textbf {\bibinfo {volume} {25}} (\bibinfo {year} {2012})}\BibitemShut {NoStop}%
\bibitem [{\citenamefont {Simonyan}\ and\ \citenamefont {Zisserman}(2014)}]{simonyan2014very}%
  \BibitemOpen
  \bibfield  {author} {\bibinfo {author} {\bibfnamefont {K.}~\bibnamefont {Simonyan}}\ and\ \bibinfo {author} {\bibfnamefont {A.}~\bibnamefont {Zisserman}},\ }\bibfield  {title} {\bibinfo {title} {Very deep convolutional networks for large-scale image recognition},\ }\href@noop {} {\bibfield  {journal} {\bibinfo  {journal} {arXiv preprint arXiv:1409.1556}\ } (\bibinfo {year} {2014})}\BibitemShut {NoStop}%
\bibitem [{\citenamefont {He}\ \emph {et~al.}(2016)\citenamefont {He}, \citenamefont {Zhang}, \citenamefont {Ren},\ and\ \citenamefont {Sun}}]{he2016deep}%
  \BibitemOpen
  \bibfield  {author} {\bibinfo {author} {\bibfnamefont {K.}~\bibnamefont {He}}, \bibinfo {author} {\bibfnamefont {X.}~\bibnamefont {Zhang}}, \bibinfo {author} {\bibfnamefont {S.}~\bibnamefont {Ren}},\ and\ \bibinfo {author} {\bibfnamefont {J.}~\bibnamefont {Sun}},\ }\bibfield  {title} {\bibinfo {title} {Deep residual learning for image recognition},\ }in\ \href@noop {} {\emph {\bibinfo {booktitle} {Proceedings of the IEEE conference on computer vision and pattern recognition}}}\ (\bibinfo {year} {2016})\ pp.\ \bibinfo {pages} {770--778}\BibitemShut {NoStop}%
\bibitem [{\citenamefont {LeCun}\ \emph {et~al.}(2015)\citenamefont {LeCun}, \citenamefont {Bengio},\ and\ \citenamefont {Hinton}}]{lecun2015deep}%
  \BibitemOpen
  \bibfield  {author} {\bibinfo {author} {\bibfnamefont {Y.}~\bibnamefont {LeCun}}, \bibinfo {author} {\bibfnamefont {Y.}~\bibnamefont {Bengio}},\ and\ \bibinfo {author} {\bibfnamefont {G.}~\bibnamefont {Hinton}},\ }\bibfield  {title} {\bibinfo {title} {Deep learning},\ }\href@noop {} {\bibfield  {journal} {\bibinfo  {journal} {nature}\ }\textbf {\bibinfo {volume} {521}},\ \bibinfo {pages} {436} (\bibinfo {year} {2015})}\BibitemShut {NoStop}%
\bibitem [{\citenamefont {Tinto}\ and\ \citenamefont {Dhurandhar}(2021)}]{Tinto}%
  \BibitemOpen
  \bibfield  {author} {\bibinfo {author} {\bibfnamefont {M.}~\bibnamefont {Tinto}}\ and\ \bibinfo {author} {\bibfnamefont {S.~V.}\ \bibnamefont {Dhurandhar}},\ }\bibfield  {title} {\bibinfo {title} {Time-delay interferometry},\ }\href {https://doi.org/10.1007/s41114-020-00029-6} {\bibfield  {journal} {\bibinfo  {journal} {Living Reviews in Relativity}\ }\textbf {\bibinfo {volume} {24}},\ \bibinfo {pages} {1} (\bibinfo {year} {2021})}\BibitemShut {NoStop}%
\bibitem [{\citenamefont {Armstrong}\ \emph {et~al.}(1999)\citenamefont {Armstrong}, \citenamefont {Estabrook},\ and\ \citenamefont {Tinto}}]{armstrong1999time}%
  \BibitemOpen
  \bibfield  {author} {\bibinfo {author} {\bibfnamefont {J.}~\bibnamefont {Armstrong}}, \bibinfo {author} {\bibfnamefont {F.}~\bibnamefont {Estabrook}},\ and\ \bibinfo {author} {\bibfnamefont {M.}~\bibnamefont {Tinto}},\ }\bibfield  {title} {\bibinfo {title} {Time-delay interferometry for space-based gravitational wave searches},\ }\href@noop {} {\bibfield  {journal} {\bibinfo  {journal} {The Astrophysical Journal}\ }\textbf {\bibinfo {volume} {527}},\ \bibinfo {pages} {814} (\bibinfo {year} {1999})}\BibitemShut {NoStop}%
\bibitem [{\citenamefont {Babak}\ \emph {et~al.}(2021)\citenamefont {Babak}, \citenamefont {Hewitson},\ and\ \citenamefont {Petiteau}}]{Babak2021}%
  \BibitemOpen
  \bibfield  {author} {\bibinfo {author} {\bibfnamefont {S.}~\bibnamefont {Babak}}, \bibinfo {author} {\bibfnamefont {M.}~\bibnamefont {Hewitson}},\ and\ \bibinfo {author} {\bibfnamefont {A.}~\bibnamefont {Petiteau}},\ }\href {http://arxiv.org/abs/2108.01167} {\bibinfo {title} {{LISA} {Sensitivity} and {SNR} {Calculations}}} (\bibinfo {year} {2021}),\ \bibinfo {note} {arXiv:2108.01167 [astro-ph, physics:gr-qc]}\BibitemShut {NoStop}%
\bibitem [{\citenamefont {Luo}\ \emph {et~al.}(2020)\citenamefont {Luo}, \citenamefont {Guo}, \citenamefont {Jin}, \citenamefont {Wu},\ and\ \citenamefont {Hu}}]{luo2020brief}%
  \BibitemOpen
  \bibfield  {author} {\bibinfo {author} {\bibfnamefont {Z.}~\bibnamefont {Luo}}, \bibinfo {author} {\bibfnamefont {Z.}~\bibnamefont {Guo}}, \bibinfo {author} {\bibfnamefont {G.}~\bibnamefont {Jin}}, \bibinfo {author} {\bibfnamefont {Y.}~\bibnamefont {Wu}},\ and\ \bibinfo {author} {\bibfnamefont {W.}~\bibnamefont {Hu}},\ }\bibfield  {title} {\bibinfo {title} {A brief analysis to taiji: Science and technology},\ }\href@noop {} {\bibfield  {journal} {\bibinfo  {journal} {Results in Physics}\ }\textbf {\bibinfo {volume} {16}},\ \bibinfo {pages} {102918} (\bibinfo {year} {2020})}\BibitemShut {NoStop}%
\bibitem [{\citenamefont {Nitz}\ \emph {et~al.}(2022)\citenamefont {Nitz}, \citenamefont {Harry}, \citenamefont {Brown}, \citenamefont {Biwer}, \citenamefont {Willis}, \citenamefont {Dal~Canton}, \citenamefont {Capano}, \citenamefont {Dent}, \citenamefont {Pekowsky}, \citenamefont {Williamson} \emph {et~al.}}]{nitz2022gwastro}%
  \BibitemOpen
  \bibfield  {author} {\bibinfo {author} {\bibfnamefont {A.}~\bibnamefont {Nitz}}, \bibinfo {author} {\bibfnamefont {I.}~\bibnamefont {Harry}}, \bibinfo {author} {\bibfnamefont {D.}~\bibnamefont {Brown}}, \bibinfo {author} {\bibfnamefont {C.~M.}\ \bibnamefont {Biwer}}, \bibinfo {author} {\bibfnamefont {J.}~\bibnamefont {Willis}}, \bibinfo {author} {\bibfnamefont {T.}~\bibnamefont {Dal~Canton}}, \bibinfo {author} {\bibfnamefont {C.}~\bibnamefont {Capano}}, \bibinfo {author} {\bibfnamefont {T.}~\bibnamefont {Dent}}, \bibinfo {author} {\bibfnamefont {L.}~\bibnamefont {Pekowsky}}, \bibinfo {author} {\bibfnamefont {A.~R.}\ \bibnamefont {Williamson}}, \emph {et~al.},\ }\bibfield  {title} {\bibinfo {title} {gwastro/pycbc: v2. 0.2 release of pycbc},\ }\href@noop {} {\bibfield  {journal} {\bibinfo  {journal} {Zenodo}\ } (\bibinfo {year} {2022})}\BibitemShut {NoStop}%
\bibitem [{\citenamefont {Boh{\'e}}\ \emph {et~al.}(2017)\citenamefont {Boh{\'e}}, \citenamefont {Shao}, \citenamefont {Taracchini}, \citenamefont {Buonanno}, \citenamefont {Babak}, \citenamefont {Harry}, \citenamefont {Hinder}, \citenamefont {Ossokine}, \citenamefont {P{\"u}rrer}, \citenamefont {Raymond} \emph {et~al.}}]{bohe2017improved}%
  \BibitemOpen
  \bibfield  {author} {\bibinfo {author} {\bibfnamefont {A.}~\bibnamefont {Boh{\'e}}}, \bibinfo {author} {\bibfnamefont {L.}~\bibnamefont {Shao}}, \bibinfo {author} {\bibfnamefont {A.}~\bibnamefont {Taracchini}}, \bibinfo {author} {\bibfnamefont {A.}~\bibnamefont {Buonanno}}, \bibinfo {author} {\bibfnamefont {S.}~\bibnamefont {Babak}}, \bibinfo {author} {\bibfnamefont {I.~W.}\ \bibnamefont {Harry}}, \bibinfo {author} {\bibfnamefont {I.}~\bibnamefont {Hinder}}, \bibinfo {author} {\bibfnamefont {S.}~\bibnamefont {Ossokine}}, \bibinfo {author} {\bibfnamefont {M.}~\bibnamefont {P{\"u}rrer}}, \bibinfo {author} {\bibfnamefont {V.}~\bibnamefont {Raymond}}, \emph {et~al.},\ }\bibfield  {title} {\bibinfo {title} {Improved effective-one-body model of spinning, nonprecessing binary black holes for the era of gravitational-wave astrophysics with advanced detectors},\ }\href@noop {} {\bibfield  {journal} {\bibinfo  {journal} {Physical Review D}\ }\textbf {\bibinfo {volume} {95}},\ \bibinfo {pages} {044028} (\bibinfo
  {year} {2017})}\BibitemShut {NoStop}%
\bibitem [{\citenamefont {Wang}\ \emph {et~al.}(2020{\natexlab{b}})\citenamefont {Wang}, \citenamefont {Ni}, \citenamefont {Han}, \citenamefont {Yang},\ and\ \citenamefont {Zhong}}]{wang2020numerical}%
  \BibitemOpen
  \bibfield  {author} {\bibinfo {author} {\bibfnamefont {G.}~\bibnamefont {Wang}}, \bibinfo {author} {\bibfnamefont {W.-T.}\ \bibnamefont {Ni}}, \bibinfo {author} {\bibfnamefont {W.-B.}\ \bibnamefont {Han}}, \bibinfo {author} {\bibfnamefont {S.-C.}\ \bibnamefont {Yang}},\ and\ \bibinfo {author} {\bibfnamefont {X.-Y.}\ \bibnamefont {Zhong}},\ }\bibfield  {title} {\bibinfo {title} {Numerical simulation of sky localization for lisa-taiji joint observation},\ }\href@noop {} {\bibfield  {journal} {\bibinfo  {journal} {Physical Review D}\ }\textbf {\bibinfo {volume} {102}},\ \bibinfo {pages} {024089} (\bibinfo {year} {2020}{\natexlab{b}})}\BibitemShut {NoStop}%
\bibitem [{\citenamefont {Wang}\ and\ \citenamefont {Ni}(2023)}]{wang2023revisiting}%
  \BibitemOpen
  \bibfield  {author} {\bibinfo {author} {\bibfnamefont {G.}~\bibnamefont {Wang}}\ and\ \bibinfo {author} {\bibfnamefont {W.-T.}\ \bibnamefont {Ni}},\ }\bibfield  {title} {\bibinfo {title} {Revisiting time delay interferometry for unequal-arm lisa and taiji},\ }\href@noop {} {\bibfield  {journal} {\bibinfo  {journal} {Physica Scripta}\ }\textbf {\bibinfo {volume} {98}},\ \bibinfo {pages} {075005} (\bibinfo {year} {2023})}\BibitemShut {NoStop}%
\bibitem [{\citenamefont {Diakogiannis}\ \emph {et~al.}(2020)\citenamefont {Diakogiannis}, \citenamefont {Waldner},\ and\ \citenamefont {Caccetta}}]{diakogiannis2020looking}%
  \BibitemOpen
  \bibfield  {author} {\bibinfo {author} {\bibfnamefont {F.~I.}\ \bibnamefont {Diakogiannis}}, \bibinfo {author} {\bibfnamefont {F.}~\bibnamefont {Waldner}},\ and\ \bibinfo {author} {\bibfnamefont {P.}~\bibnamefont {Caccetta}},\ }\bibfield  {title} {\bibinfo {title} {Looking for change? roll the dice and demand attention},\ }\href@noop {} {\bibfield  {journal} {\bibinfo  {journal} {arXiv preprint arXiv:2009.02062}\ } (\bibinfo {year} {2020})}\BibitemShut {NoStop}%
\end{thebibliography}%

\end{document}